# Efficient Techniques for Mining Spatial Databases

By

Mohamed A. El-Zawawy

B. Computer Science, Math. Department,

Faculty of Science, Cairo University, 1999

A THESIS SUBMITTED IN PARTIAL FULFILMENT

OF THE REQUYIREMENTS FOR THE DEGREE OF

MASTER OF SIENCE

in the Math. Department,

Faculty of Science, Cairo University

Mohamed A. El-Zawawy 2002

Math. Department, Faculty of Science,

Cairo University

Supervisors

**Prof. Dr. Laila F. Abdelall**  **Dr. Mohamed E. El-Sharkawi**

Dept. of Mathematics          Dept. of Information Systems
Faculty of Science            Faculty of Computers & Information
Cairo University              Cairo University

# APPROVAL

**Name:** Mohamed A. El-Zawawy

**Degree:** Master of Science

**Title of thesis:** Efficient Techniques for Mining Spatial Databases

**Supervisors:** Dr. Laelia F. Abdel-aal

-------------------------------------------------------------------

Dr. Mohamed E. El-Sharkawi

# Abstract

Clustering is one of the major tasks in data mining. In the last few years, Clustering of spatial data has received a lot of research attention. Spatial databases are components of many advanced information systems like geographic information systems VLSI design systems. In this thesis, we introduce several efficient algorithms for clustering spatial data.

First, we present a grid-based clustering algorithm that has several advantages and comparable performance to the well known efficient clustering algorithm. The algorithm has several advantages. The algorithm does not require many input parameters. It requires only three parameters, the number of the points in the data space, the number of the cells in the grid and a percentage. The number of the cells in the grid reflects the accuracy that should be achieved by the algorithm. The algorithm is capable of discovering clusters of arbitrary shapes. The computational complexity of the algorithm is comparable to the complexity of the most efficient clustering algorithm. The algorithm has been implemented and tested against different ranges of database sizes. The performance results show that the running time of the algorithm is superior to the most well known algorithms (CLARANS [23]). The results show also that the performance of the algorithm do not degrade as the number of the data points increases.

The second contribution of the thesis is extending the proposed clustering algorithm to handle obstacles. Most of the spatial clustering algorithms do not take into consideration the existence of physical obstacles in the real world. Such obstacles include natural obstacles like rivers and mountains and artificial obstacles like mine fields. We propose two algorithms that consider clustering with the existence of obstacles. We first propose a simple algorithm that handles obstacles and then the algorithm is refined to produce more fine clusters. The algorithms have the advantages of the grid-based algorithm proposed in the thesis. Moreover the algorithm handle obstacles of arbitrary geometric shapes. The second algorithm is implemented and tested against different sizes of data and obstacles of different geometric shapes. The run time characteristics of the algorithm are compared with those of COD-CKLARANS [29], an extension of CLARANS algorithm [23] to handle obstacles. The results show that the proposed algorithm out performs COD-CLARANS in general and specially when the number of the data points is extremely large.

Spatial databases contain spatial-related information such databases include geographic (map) databases. Spatial data mining is the discovery of interesting characteristics and patterns that may exist in large spatial databases. Clustering, in spatial data mining, aims at grouping a set of objects into classes or clusters such that objects within a cluster have high similarity among each other, but are dissimilar to objects in other clusters. Many clustering methods have been developed. Most of these algorithms, however, dose not allow users to specify real life constraints such as the existence of physical obstacles, like mountains and rivers. Existence of such obstacles could substantially affect the result of a clustering algorithm.

In this thesis, we proposed an efficient algorithm for spatial clustering of large spatial databases. The algorithm overcomes the problems of the previous work. The algorithm divides the spatial area into rectangular cells and labels each cell as *dense* (contains relatively large number of points) or *non-dense*. The algorithm finds all the maximal, connected, and dense regions that form the clusters by a breadth-first search and determine a center for each region.

We also implemented the proposed algorithm and compare it with CLARANS algorithm [NH94]. The experiments showed that our algorithm is more superior in both running time and accuracy of the results.

The second contribution of this work is introducing efficient algorithms of spatial clustering with presence of obstacles. In [THH01], the problem of clustering with presence of obstacles is introduced.

The proposed algorithms are called as *CPO-WCC (Clustering in Presence of Obstacles with Computed number of cells)*, and *CPO-WFC (Clustering in Presence of Obstacles with Fixed number of cells)*.

The proposed algorithms, *CPO-WCC*, and *CPO-WFC* have several advantages over other work in [THH01]. They handle outliers or noise. Outliers refer to spatial objects, which are not contained in any cluster and should be discarded during the mining process. Second, they do not use any randomized search. Third Instead of specifying the number of desired clusters beforehand, They find the natural number of clusters in the area. Finally, when the data is updated, we do not need to re-compute information for all cells in the grid. Instead, incremental update is performed.

We also implemented *CPO-WFC* and compare it with COD-CLARANS algorithm [NHH01]. The experiments showed that *CPO-WFC* is more superior in both running time and accuracy of the results.

*To my family*

# Acknowledgments

I would like to express my thanks to my supervisors professor Dr. Laila Abdelall and Dr. Mohamed El-Sharkawi, for allowing me to work under their supervision. They have been supportive and confident in my ability, and their guidance was invaluable to me during the master study. I would like to thank Dr. El-Sharkawi for proof reading my thesis. And to him I would like to say without your patience and support, I would have never finished the thesis.

I would like to take the chance to express my appreciation to my family. Their continuous love and support gave me the strength of pursuing my goal.

I would like to thank Dr Aboel-ela Afefe for his valuable comments and suggestions. Thanks go to all friends I made during my master study.

# Contents

# List of Tables

# List of Figures

# Chapter 1

# Introduction

Clustering is a descriptive task that seeks to identify homogeneous groups of objects based on the values of their attributes (dimensions) [18] [21]. Clustering techniques have been studied extensively in statistics [3], pattern recognition [10] [12], and machine learning [9] [22]. Recently, The Clustering techniques have used the database area. Recent work in the database community includes density-based methods, hierarchal-based methods, partition-based methods, grid-based methods, and constrained-based methods.

## 1.1 Spatial Clustering

Spatial databases contain spatial-related information such databases include geographic (map) databases, VLSI chip design databases, and medical and satellite image databases. Spatial data mining is the discovery of interesting characteristics and patterns that may exist in large spatial databases. Clustering, in spatial data mining, aims at grouping a set of objects into classes or clusters such that objects within a cluster have high similarity among each other, but are dissimilar to objects in other clusters. Spatial clustering is used in many applications such as seismology (grouping earthquakes clustered along seismic faults) and geographic information systems (GIS).

In the last few years, clustering in spatial databases has received a lot of research attention. Many algorithms have been proposed to perform spatial clustering. These algorithms can be classified into five categories [16]: density-based methods, hierarchal-based methods, partition-based methods, grid-based methods, and constrained-based methods.

Density-based methods [6, 7, 25] view clusters as dense regions of objects that are separated by low density regions in the data space. Density-based methods have the advantage of discovering the arbitrary shaped clusters.

Hierarchal-based methods [4, 19, 13] put the data in a tree of clusters. The hierarchal-based clusters are classified into agglomerative and divisive hierarchical clustering, depending on whether the decomposition is formed in a bottom-up or top-down manner.

A partitioning algorithm [15, 23]divides *n* objects, which we want to cluster, into *k* partitions, where each partition represents a cluster and *k* is a given parameter. Partitioning-based algorithms form the clusters by optimizing an objective criterion, similarity function, such as distance.

Grid-based clustering algorithms [31, 26, 2] summarize the data space into a finite number of cells that form a grid structure on which all of the operations for clustering are performed. These methods always have fast processing time, which typically independent of the number of the data objects, but dependent on the number of the cells in each dimension in the summarized space.

Most of these clustering algorithms does not allow users to specify real life constraints such as physical obstacles. A constrained-based algorithm [30, 29] allows users to specify constraints on the data set being clustered.

## 1.2  Requirements

The requirements of spatial clustering algorithms which raised by application of large spatial databases are the following: minimal knowledge of the domain to determine input parameters, discovering of clusters with arbitrary shapes, and acceptable efficiency in handling large databases. Most of the proposed algorithms do not satisfy all the three requirements.

Many clustering methods have been developed. Most of these algorithms, however, dose not allow users to specify real life constraints such as the existence of physical obstacles, like mountains and rivers. Existence of such obstacles could substantially affect the result of a clustering algorithm. For example, consider a telephone-company that wishes to locate a suitable number of telephone cabinets in the area shown in Figure 1.1 to serve the customers who are represented by points in the figure. There are natural obstacles in the area and they should not be ignored. Ignoring these obstacles will result in clusters like those in Figure 1.2, which are obviously inappropriate. Cluster $cl_1$ is split by a river, and customers on one side of the river have to travel a long way to reach the telephone cabinet at the other side. Thus the ability to handle such real life constraints in a clustering algorithm is important.

In a recent paper, [29], the problem of clustering with existence of obstacles is introduced, and an algorithm, COD-CLARANS, to solve the problem was introduced. Although COD-CLARANS generates good clustering results, it has several major problems. First, since the randomized search is used in the algorithm to determine initial centers for the clusters and then to refine those centers, the quality of the results cannot be guaranteed when the number of points, *N*, is large. Second, COD-CLARANS takes as an input the number of the desired clusters and another integer, which determine the number of maximum tries to refine center, but both numbers are generally unknowns in realistic applications. Third, COD-CLARANS can't handle outliers. Forth, when data is updated, we need to run the algorithm from scratch.

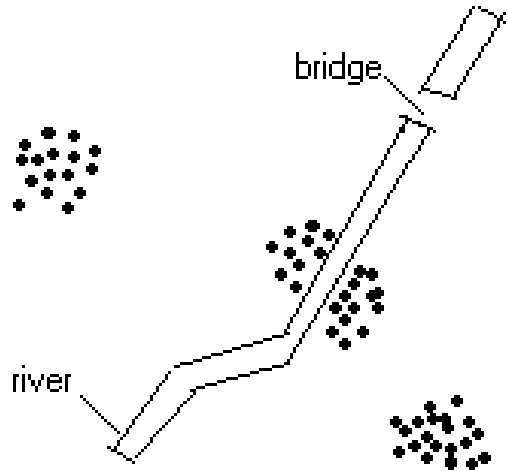

Figure 1.1 :Customers' locations and obstacles

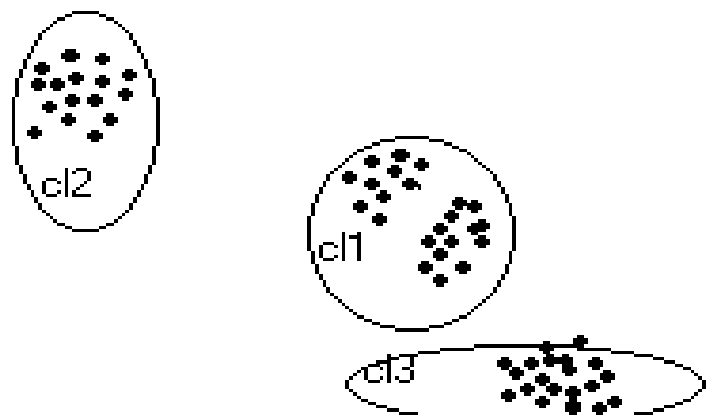

Figure 1.2: Clusters formed when ignoring obstacles

## 1.3 Contributions

1. In this thesis, we propose an efficient algorithm for spatial clustering of large databases that we call *SCLD (Spatial Clustering in Large Databases)*, which overcomes the disadvantages of the previous work. The algorithm divides the spatial area into rectangular cells and labels each cell as *dense* (contains relatively large number of points) or *non-dense*. The algorithm finds all the maximal, connected, dense regions that form the clusters by a breadth-first search and determine a center for each region. We also implement the SCLD algorithm and compare it with one of the most robust clustering algorithms.

We believe that The *SCLD* algorithm has the following advantages over the previous work:
1. It handles outliers or noise. Outliers refer to spatial objects, which are not contained in any clusters and should be discarded during the mining process. *CLARANS,* for example, can not handle outliers.
2. It requires only two parameters, the number of the cells in the grid structure a percentage *h* used as in Definition 2. Instead of specifying the number of the desired clusters beforehand as input (as in *CLARANS* ), the *SCLD* algorithm finds the natural number of clusters in the spatial area.
3. When the data is updated, we do not need to re-compute all the information in the grid structure. Instead, we can do an incremental update. This is done by re-compute the information(the number of points and the mean point ) of cells that included the update. The re-compute the clusters by a breadth-first search on the cells of the grid structure.
4. It discovers clusters of arbitrary shape and it is efficient even for large spatial databases. However, *CLARANS* failed to discover clusters of arbitrary shapes. Furthermore *CLARANS* can not deal with noise.
5. Its computational complexity is much less than that of *CLARANS*.

2. In this thesis, we propose two different efficient spatial clustering algorithms, *CPO-WCC*, *CPO-WFC* , which consider the presence of obstacles. We also implement the *CPO-WFC* algorithm and compare it with the only clustering algorithm (COD-CLARANS) that consider the presence of obstacles.

The proposed algorithms(*CPO-WCC* and *CPO-WFC*) have several advantages over COD-CLARANS [29].

1. They handle outliers or noise. Outliers refer to spatial objects, which are not contained in any cluster and should be discarded during the mining process.

2. They dose not use any randomized search.

3. Instead of specifying the number of desired clusters beforehand, They find the natural number of clusters in the area.

4. When the data is updated, we do not need to re-compute information for all cells in the grid. Instead, incremental update is performed.

## 1.4  Thesis Organization

Our thesis is comprised of five chapters. In Chapter2, we survey methods in spatial clustering. In Chapter 3, we propose a clustering algorithm for very large spatial databases that satisfy the aforementioned requirements and we also show our experimental results. In Chapter 4, we propose efficient spatial clustering techniques, which consider the presence of obstacles and we also show our expermental results. In Chapter 5, we summarize the technical contributions of this thesis and present possible directions for future research.

# Chapter 2

# Related Work

In this chapter, we will survey the current researches in spatial clustering methods. The spatial clustering algorithms can be classified into five categories: density-based methods, hierarchal-based methods, partition-based methods, grid-based methods, constrained-based methods. In this chapter, we will introduce the most well known algorithms in each category and discuss their strenuous and weakness.

## 2.1 Density-based methods

The density-based methods view the clusters as dense regions of objects that are separated by low density regions in the data space. The density-base methods have the advantage of discovering the arbitrary shaped clusters. In this section, we will introduce the well-known algorithms under this category.

### LOF: Identifying Density-based local outliers

LOF (local outlier factor) [BKNS00] is a density-based method that finds the rare instances or the outliers. [BKNS00] assigns to each object in a database a degree of being an outlier that is called the local outlier factor (*LOF*). This degree is local in that it depends on how isolated the object is with respect to the surrounding neighborhood.

[BKNS00] introduces the following new terminologies to define the local outlier factor.

1. For any positive integer *k*, the **k-distance** of object *p*, denoted as *k - distance*(p), is defined as the distance *d(p,o)* between *p* and an object $o \in D$ such that:

    A. for at least *k* objects $o' \in D \setminus \{ p \}$ it holds that $d(p,o') \leq d(p,o)$, and

    B. for at most *k-1* objects $o' \in D \setminus \{ p \}$ t holds that $d(p,o') < d(p,o)$.

2. Given the *k-distance* of *p*, the **k-distance neighborhood** of *p* contains every object whose distance from *p* is not greater than the *k-distance*, i.e. $N_{k-distance(p)}(p) = \{ q \in D \setminus \{ p \} / d(p,q) \leq k - dis\tan ce(p) \}$.

3. Let *k* be a natural number. The **reachability distance** of object *p* with respect to object *o* is defined as *reach-dist$_k$ (p, o)* = max { *k-distance(o), d(p, o)*}.

4. The **local reachability density** of *p* is defined as

$$lrd_{MinPts}(p) = 1 \Bigg/ \left( \frac{\sum_{o \in N_{MinPts}(p)} reach-dist_{MinPts}(p,o)}{|N_{MinPts}(p)|} \right)$$

5. The **(local) outlier factor** of *p* is defined as

$$LOF_{MinPts}(p) = \frac{\sum_{o \in N_{MinPts}(p)} \frac{lrd_{MinPts}(o)}{lrd_{MinPts}(p)}}{|N_{MinPts}(p)|}$$

*LOF(p)* is simply a function of the reachability distances in *p*'s direct neighborhood relative to those in *p*'s indirect neighborhood.

[BKNS00] proves that the *local outlier factor* (*LOF*) have the following properties.

1. For most objects *p* in a cluster, the *LOF* of *p* in a cluster is approximately equal to 1.

2. Let *p* be an object from the database *D*, and $1 \leq MinPts \leq |D|$.

Then, it is the case that

$$\frac{direct_{min}(p)}{indirect_{max}(p)} \leq LOF(p) \leq \frac{direct_{max}(p)}{indirect_{min}(p)}$$

where $direct_{min}(p) = \min \{ reach\text{-}dist(p, q) \mid q \in N_{MinPts}(p) \}$ and

$indirect_{min}(p) = \min \{ reach\text{-}dist(q, o) \mid q \in N_{MinPts}(p)$ and $o \in N_{MinPts}(q) \}$.

3. For an object $p$ which is not located deep inside a cluster, but whose *MinPts*-nearest neighbors all belong to the same cluster the bounds on *LOF* as predicted in step 2 are tight.

4. Let $p$ be an object from the database $D$, $1 \leq MinPts \leq |D|$, and $C_1, C_2, ..., C_n$ be a partition of $N_{MinPts}(p)$, i.e. $N_{MinPts}(p) = C_1 \cup C_2 \cup ... \cup C_n \cup \{p\}$ with $C_i \cap C_j = \emptyset$, $C_i \neq \emptyset$ for $1 \leq i,j \leq n$, $i \neq j$.

Furthermore, let $\xi_i = |C_i|/|N_{MinPts}(p)|$ be the percentage of objects in $p$'s neighborhood, which are also in $C_i$. Let the notions $direct^i_{min}(p), direct^i_{max}(p), indrect^i_{min}(p),$ and $indirect^i_{max}(p)$ be defined analogously to $direct_{min}(p), direct_{max}(p), indrect_{min}(p),$ and $indirect_{max}(p)$ but restricted to the set $C_i$ (e.g., denotes the minimum reachability distance between $p$ and a *MinPts*-nearest neighbor of $p$ in the set $C_i$).

Then, it holds that (a)

$$LOF(p) \geq \left( \sum_{i=1}^{n} \xi_i \cdot direct^i_{min}(p) \right) \left( \sum_{i=1}^{n} \frac{\xi_i}{indirect^i_{max}(p)} \right)$$

and (b)

$$LOF(p) \leq \left( \sum_{i=1}^{n} \xi_i \cdot direct^i_{max}(p) \right) \cdot \left( \sum_{i=1}^{n} \frac{\xi_i}{indirect^i_{min}(p)} \right)$$

6. The *LOF* neither decreases nor increases monotonically with respect to *MinPts*. Because the *LOF* can go up and down, [BKNS00] propose as a heuristic that they use a range of *MinPts* values. [BKNS00] use *MinPtsLB* and *MinPtsUB* to denote the "lower bound" and the "upper bound" of the range. Experiments show that *MinPtsLB* should be at least 10 to remove un-wanted statistical fluctuations. The second guideline they provide for picking *MinPtsUB* is the maximum number of "close by" objects that can potentially be local outliers.

[BKNS00] produces good results but only for the appropriate choices of the parameter *MinPts*.

**Density Biased Sampling**

Density Biased sampling [PF00] is an improved method for data mining and clustering. Many clustering methods use a *p* uniform sample (a sample in which each element has probability *p* of bring selected). Those algorithms select a sample from the database and cluster them. Provided that the sample was representative of the data, the clustering is generalized to the entire data set. Figure 2.1 illustrates why uniform sampling is not necessary a good choice. The example in Figure 1 contains 4 clusters. Clusters A and B each contains 9,900 points while clusters C and D each contains only 100 points. A 1% sample of the data set would be expected to draw around 99 points from each of A and B and a single point from each of C and D. So the points from C and D will likely be treated as noise by the clustering algorithm. That is we expect that clusters C and D will be completely missed.

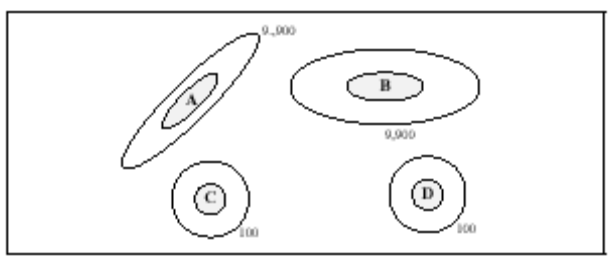

Figure 2.1: Four clusters with skewed sizes

Density Biased Sampling [PF00] introduces a new sampling technique an efficient algorithm that improves on uniform sampling when cluster sizes are skewed.

[PF00] defines a probability function and a corresponding weighting of the sample points that satisfies:
1. Within a group points are selected uniformly.
2. The sample is density preserving.
3. The sample is biased by the group size.
4. Expected sample size is *M*.

By the density preserving [PF00] means that the expected sum of the weights of the points for each group is proportional to the group's size. To satisfy criterion 1,[PF00] defines P(selecting point $x \mid x$ in group $i$ )= $f(n_i)$, where $n_i$ is the size of group $i$. Also each point from the group is assigned equal weight $w(n_i) = 1/f(n_i)$. So the expected weight of the points in group $i$ is:

$$\sum_{j=1}^{n_i} P(point\ x_i) \cdot w(n_i) = \sum_{j=1}^{n_i} f(n_i) \cdot 1/f(n_i) = n_i.$$

To biases the sample by group size, [PF00] defines $f(n_i) = \dfrac{\alpha}{n_i^e}$, for any constant $e$. $\alpha$ is defined such that the expected sample size is $M$:

$$E(sample\ size) = \sum_{i=1}^{g} E(size\ of\ group\ i)$$

$$M = \sum_{i=1}^{g} n_i \cdot f(n_i) = \sum_{i=1}^{g} n_i \cdot \frac{\alpha}{n_i^e}$$

$$\Rightarrow \alpha = \frac{M}{\sum_{i=1}^{g} n_i^{1-e}}$$

To partition the data into groups those are required by Density Biased Sampling and assign probabilities and weights that satisfy the four conditions, [PF00] works as follows. Numerical attributes are divided into G bins and categorical attributes have a bin for each category. The space is divided into bins by placing a *d*-dimensional grid over the data. [PF00] employees a hashing based approach to create an array of bin counts. This array is called *n* (then *n[i]* corresponds to $n_i$ above) and has *H* entries (correspond to the number of the groups) indexed from 0. To index into this array, [PF00] uses a hash function from the bin label to array index (see Figure 2.2). Finally the

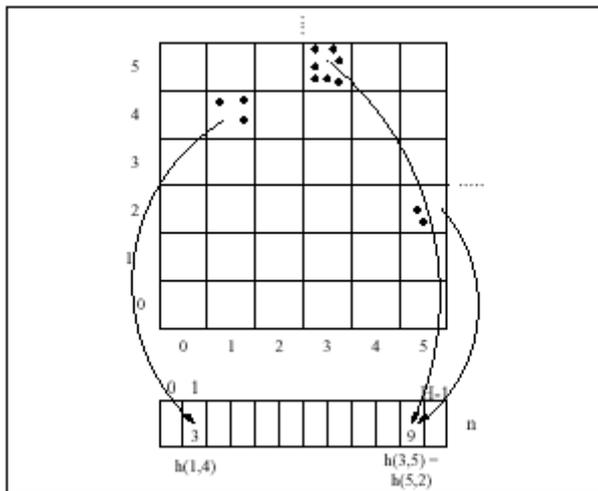

Figure 2.2 Density approximation by hashing

algorithm does the following steps:

1. For each input vector $x$, $n[h(x)] = n[h(x)]+1$.
2. Compute $\alpha$.
3. For each input vector $x$, with probability $\alpha/n[h(x)]^e$, output $<n[h(x)]^e/\alpha, x>$.

[PF00] eliminates the second pass over the data in the final step by built the densities and the sample in parallel. This is done based on the observation that, If when the data is restricted to the first $j$ records, the probability of outputting some record $x$ is $P_j$. Then for $j \leq j', p_j \geq p_{j'}$.

The one pass hash approximation to density biased sampling works as follows. A buffer of points that have some chance of being in the sample will be maintained. The buffer contains elements $\{<P_i, x_i>\}$ to indicate that $x_i$ was added to the buffer with probability $P_i$. Suppose that at some later point, $x_i$ would have probability $P_i'$ of being output. [PF00] can convert the current output buffer into a buffer that is a density biased sample of the currently processed data. The lemma tells us that $P_i' \leq P_i$ and consequently [PF00] will never erroneously discard a point due to an underestimate of its probability. If [PF00] keep $<P_i', x_i>$ in the buffer with probability $P_i'/P_i$ (otherwise, remove this entry from the buffer), then $x_i$ is in the buffer with probability $P_i'$. The weight of a point is just $1/P_i'$ which means that we can output the weighted sample from the reduced buffer.

[PF00] is robust to poor parameter choices.

**DBSCAN**

DBSCAN (Density-Based Spatial Clustering of Application with Noise)[EKSX96] is a density-based clustering algorithm.

Given a database *D* and the user-defined parameters *Eps, MinPts,* DBSCAN introduces the following new terminologies:

1. The **Eps-neighborhood** of a point *p*, denoted by $N_{Eps}(p)$, is defined by $N_{Eps} = \{q \in D \mid dist(p,q) \leq Eps\}$.

2. A point *p* is **directly density-reachable** from *q* wrt. *Eps* and *MinPts* if

    1) $p \in N_{Eps}(q)$ and

    2) $|N_{Eps}(q)| \geq MinPts$ (core point condition).

3. A point *p* is **density-reachable** from *q* wrt. *Eps* and *MinPts* if there is a chain of points $p_1,...,p_n, p_1 = q, p_n = p$ such that $p_{i+1}$ is directly density-reachable from $p_i$.

4. A point *p* is **density-connected** to a point *q* wrt. *Eps* and *MinPts* if there a point *o* such that both *p* and *q* are density reachable from *o* wrt. *Eps* and *MinPts*.

5. A **cluster** *C* wrt. *Eps* and *MinPts* is a non-empty subset of *D* satisfying the following conditions:

    *1)* $\forall p,q$: if $p \in C$ and *q* is density reachable from *p* wrt. *Eps* and *MinPts,* then $q \in C$.

    *2)* $\forall p,q \in C$: *p* is density-connected to *q* wrt. *Eps* and *MinPts*.

6. Let $C_1,...,C_k$ be the clusters of the database *D* wrt. $Eps_i$ and $MinPts_i, i = 1,...,k$, Then we define the **noise** as the set of points in the database *D* not belonging to any cluster $C_i$, i.e. **noise** $= \{p \in D \mid \forall i : p \notin C_i\}$

Figure 2.3 illustrates the definitions on a sample database of 2-dimensional points from a vector space.

DBSCAN works as follows. Starting with an arbitrary point, *p*, the algorithm retrieves all points that are density-reachable from *p* wrt. *Eps* and *MinPts*. The retrieval of *density-reachable* objects is performed by iteratively collecting *directly density-reachable* objects. If *p* is a core point (i.e. $|N_{Eps}(p)| \geq MinPts$), then *p* and all points that are density reachable are collected in one cluster. If *p* isn't a core point, then p is considered as an outlier and discarded later if it isn't assigned to any cluster. The algorithm terminates when no new points can be assigned to any cluster.

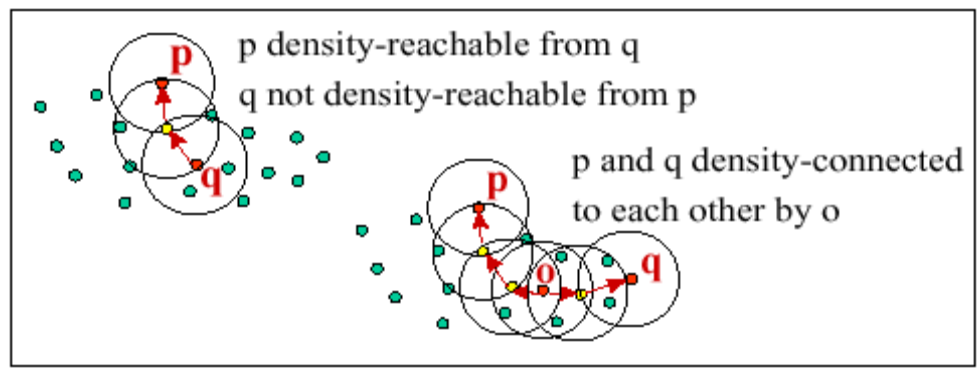

Figure 2.3: Density reachibility and connectivity.

The average complexity of the DBSCAN is *O(nlogn)*. This is because DBSCAN supports region queries by using the spatial data type $R^*$-tree.

DBSCAN has the advantage of discovering clusters that have arbitrary shapes. On the other hand, DBSCAN may merge two clusters that are sufficiently close to each other.

**OPTICS**

OPTICS (Ordering Points To Identify Clustering Structure) [ABKS99] is an extension to BDSCAN algorithm. In the following we use the symbol $\varepsilon$ corresponding to *Eps* in the above definitions of DBSCAN. OPTICS created to overcome the DBSCAN problem arises by observing that for a constant *MinPts*-value, density-based clusters with respect to a higher density (i.e. a lower value for $\varepsilon$) are contained in density-connected sets with respect to a lower density (i.e. a higher value for $\varepsilon$). Figure 2.4 depicts this fact, where $C_1$ and $C_2$ are density-based clusters with respect to $\varepsilon_2 < \varepsilon_1$ and C is a density-based cluster with respect to $\varepsilon_1$ containing the sets $C_1$ and $C_2$. So the DBSCAN problem is that DBSCAN is very sensitive to the input parameters $\varepsilon$ and *MinPts* and choosing these parameters can be difficult.

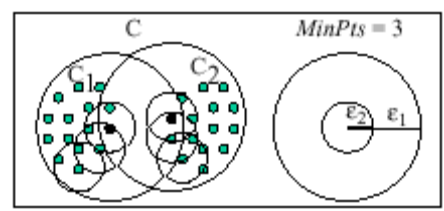

Figure 2.4: illustration of "nested" density-based clusters

OPTICS introduces the following new terminologies:
1. Let *p* be an object from a database *D*, let $\varepsilon$ be a distance value, let $N_\varepsilon(p)$ be the $\varepsilon$-neighborhood of *p*, let *MinPts* be a natural number and let *MinPts-distance(p)* be the distance from *p* to its *MinPts'* neighbor. Then the **core-distance** of *p* is defined as

$$core-distance_{\varepsilon, MinPts}(p) = \begin{cases} UNDEFINED, & \text{if } card(N_\varepsilon(p)) < MinPts \\ MinPts-distance(p), & otherwise \end{cases}$$

2. Let *b* and *o* be objects from a database *D*, let $N_\varepsilon(o)$ be the $\varepsilon$-neighborhood of *o*, and let *MinPts* be a natural number. Then the **reachability-distance** of *p* with respect to *o* is defined as

$$reachability\text{-}distance_{\varepsilon, MinPts}(p, o) = \begin{cases} UNDEFINED, \\ \quad \text{if } card(N_\varepsilon(o)) < MinPts \\ max(cor-distance(o), distance(o,p)), \\ \quad otherwise \end{cases}$$

Figure 2.5 illustrates the notions of *core-distance* and *reachability-distance*.

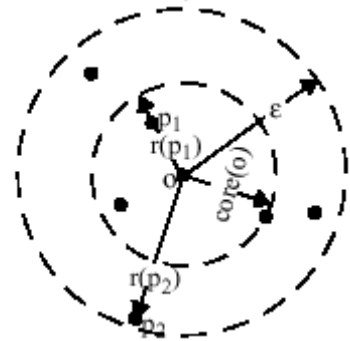

OPTICS doesn't produce a cluster of the data set explicitly; but instead creates an increasing ordering of the database objects. The objects are sorted by their *reachability-distance* to the closest core object from which they have been *directly-density reachable* with respect to given parameters $\varepsilon$, *MinPts*. So by definition, given a database containing *n* points, and two parameters $\varepsilon$ and *MinPts*, the OPTICS algorithm generates an ordering of the points

Figure 2.5: core-distance(o) reachability-distance r(p1,o), r(p2,o) for MinPts=4

$o : \{1,...,n\} \to DB$ and corresponding reachability-values $r : \{1,...,n\} \to R_{\geq 0}$.

Having the increased ordering of a database with respect to $\varepsilon$ and *MinPts,* any density-based clustering with respect to *MinPts* and a clustering-distance $\varepsilon' \leq \varepsilon$ can be easily extended by scanning the cluster-ordering and assigning cluster-memberships depending on both the reachability-distance and the core-distance of the objects.

For small data set, OPTICS represents the cluster-ordering graphically, but it uses an appropriate visualization technique for large data set. Both techniques are suitable for exploring the clustering structure, offering insights for distribution and correlation for the data. For example, Figure 2.6. depicts the reachability-plot for a simple 2-dimention data set.

For a database *D* with *n* objects, OPTICS takes an expected running time of *O(nlogn)*.

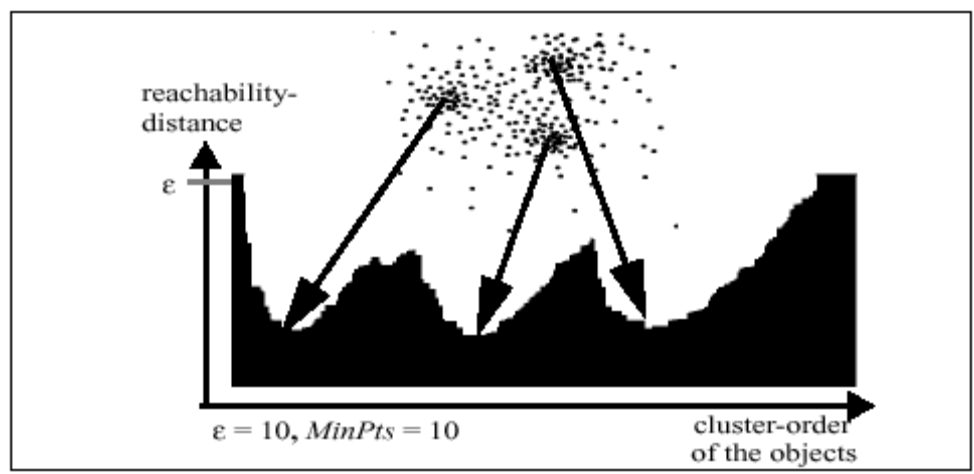

Figure 2.6: Illustration of the clusterin-ordering

**DENCLUE**

DENCLUE (DENsity CLUstEring)[HK98] is an efficient approach to clustering in large multimedia databases with noise. The approach is based on the following ideas. The influence of each data point is formally modeled using a mathematical function, which is called influence function. The influence function is a function, which describes the impact of a data point within its neighborhood. The influence function is applied to each data point. The overall density of the data space is calculated as the sum of the influence function of all data points. Clusters are determined mathematically by identifying density attractors. Density attractors are local maxima of the overall density function.

[HK98] introduces the following formal definitions for the aforementioned notions. In the following, we denote the $d$-dimensional feature space by $F^d$.

1. The *influence function* of a data object $y \in F^d$ is a function $f_B^y : F^d \to \Re_0^+$, which is defined in terms of a basic influence function $f_B$ ($f_B^y(x) = f_B(x, y)$). The *density function* is defined as the sum of the influence functions of all data points. Given $N$ data objects described by a set of feature vectors $D = \{x_1, ..., x_N\} \subset F^d$ the density function is defined as

$$f_B^D(x) = \sum_{i=1}^{N} f_B^{x_i}(x).$$

   Examples of basic influence functions are:
   
   (a) Square Wave Influence Function
   
   $$f_{square}(x, y) = \begin{cases} 0 & \text{if } d(x, y) > \sigma \\ 1 & \text{otherwise} \end{cases}$$
   
   (b) Gaussian influence function
   
   $$f_{Gauss}^D = e^{-\frac{d(x,y)^2}{2\sigma^2}}$$

2. The *gradient* of a function $f_B^D$ is defined as

$$\nabla f_B^D(x) = \sum_{i=1}^{N} (x_i - x) \cdot f_B^{x_i}(x).$$

3. A point $x^* \in F^d$ s called a *density-attractor* for a given influence function, iff $x^*$ is a local maximum for the density function $f_B^D$.

   A point $x \in F^d$ is density-attracted to a density attractor $x^*$, iff $\exists k \in N : d(x^k, x^*) \leq \varepsilon$ with $x^0 = x$, $x^i = x^{i-1} + \delta \cdot \frac{\nabla f_B^D(x^{i-1})}{\|\nabla f_B^D(x^{i-1})\|}$.

4. *A center-defined cluster* (wrt. $\sigma, \xi$) for a density-attractor $x^*$ is a subset $C \subseteq D$, with $x \in C$ being density-attracted by $x^*$ and $f_B^D(x^*) \geq \xi$.

Points $x \in D$ are called **outliers** if they are density-attracted by a local maximum $x_0^*$ with $f_B^D(x_0^*) < \xi$.

5. An **arbitrary-shape cluster** (wrt. $\sigma, \xi$) for the set of density attractors X is a subset $C \subseteq D$, where

    (a) $\forall x \in C \exists x^* \in X : f_B^D(x^*) \geq \xi$, $x$ is a density-attracted to $x^*$ and

    (b) $\forall x_1^*, x_2^* \in X : \exists$ a path $P \subset F^d$ from $x_1^*$ to $x_2^*$ with $\forall p \in P : f_B^D(p) \geq \xi$.

Given that the overall density function is differentiable at any point, the density-attractors can be calculated efficiently by a hill climbing procedure, which is guided by the gradient of the overall density function.

The algorithm DENCLUE is an efficient implementation of the above idea. Science most of the data points don't actually contribute to the density function at a point, DENCLUE use a local density function which considers only the data points, which actually contribute to the overall density function.

The DENCLUE algorithm consists of two steps. The first step is pre-clustering step in which a map of the relevant portion of the data space is constructed. The map is used to speed up the calculation of density function, which requires to efficiently access the neighboring portions of the data space. The second step is the actual clustering step in which the algorithm finds the density-attractors and the corresponding density attracted points.

DENCLUE has the following advantages. First, it has a firm mathematical basis. Second, it has a good clustering properties in data sets with large amount of noise. Third, it allows a compact mathematical description of arbitrary shaped clusters in high dimensional data sets. Finally, it is significantly faster than existing algorithms.

Similar to DBSCAN, DENCLUE is sensitive to the input parameters $\sigma, \xi$.

**Data Bubbles**

Data Bubbles [BKKS01] is a quality preserving performance boosting for hierarchal clustering. [BKKS01] scales hierarchical clustering methods (such as OPTICS) to extremely large databases.

The naïve application of a hierarchical clustering algorithm to a compressed data set (such as a random sample from the database) associated with the following problems.
   1. The clustering structure of the database is distorted. [BKKS01] calls this problem *structural distortions*.
   2. The sizes of clusters are distorted. [BKKS01] calls this problem *size distortions*.
   3. We don't have direct clustering information about the database objects. [BKKS01] calls this problem *lost objects*.

To solve the above problems, [BKKS01] introduces a new concept for a compressed data item which is called Data Bubbles. [BKKS01] also introduces the necessary terminologies to use Data Bubbles in OPTICS instead of the previous concepts of compressed data items. These new terminologies are the following.
   1. Let $X=\{X_i\} 1 \leq i \leq n$ be a set of *n* objects. Then, the **Data Bubble** *B* w.r.t. *X* is defined as a tuple $B_X = (rep, n, extent, nnDist)$, where
      - *rep* is a representative object for *X (w*hich may or may not be an element of *X)*;
      - *n* is the number of objects in *X*;
      - *extent* is a real number such that "most" objects of *X* are located within a radius" *extend* around *rep*;
      - *nnDist* is a function denoting the estimated average *k*-nearest neighbor distances within the set of objects *X* for some values *k, k=1, ..., k = MinPt*s. A particular expected *k*nn-distance in $B_X$ is denoted by $nnDist(k, B_X)$.

   2. Let B=($rep_B$, $n_B$, $e_B$, $nnDist_B$) and C=($rep_C$, $n_C$, $e_C$, $nnDist_C$) be two Data Bubbles.
      Then, the **distance** between *B* and *C* is defined as *dist*(B, C)=

$$\begin{cases} 0 & \text{if } B = C \\ dist(rep_B, rep_C) - (e_B + e_C) + nnDist(1, B) + nnDist(1, C) \\ & \text{if } dist(rep_B, rep_C) - (e_1 + e_2) \geq 0 \\ max(nnDist(1, B), nnDist(1, C)) & \text{otherwise} \end{cases}$$

3. Let B=($rep_B$, $n_B$, $e_B$, $nnDist_B$) be a Data Bubble, let $\varepsilon$ be a distance value, let *MinPts* be a natural number and let $N = \{X \mid dist(B, X) \leq \varepsilon\}$. Then, the ***core-distance*** of B is defined as

$$\text{core-dist}_{\varepsilon, MinPts}(B) = \begin{cases} \infty & \text{if } \left( \sum_{X = (r, n, e, d) \in N} n \right) < MinPts \\ dist(B, C) + nnDist(k, C) & \text{otherwise} \end{cases},$$

where C and k are given as follows: $C \in N$ has maximal $dist(B, C)$ such that

$$\sum_{\substack{X \in N \\ dist(B, X) < dist(B, C)}} n < MinPts, \text{ and } k = MinPts - \sum_{\substack{X \in N \\ dist(B, X) < dist(B, C)}} n \quad .$$

4. Let B=($rep_B$, $n_B$, $e_B$, $nnDist_B$) and C=($rep_C$, $n_C$, $e_C$, $nnDist_C$) be two Data Bubble, let $\varepsilon$ be a distance value, let *MinPts* be a natural number, and let $B \in N_C$, where $N_C = \{X \mid dist(C, X) \leq e\}$.

   Then, the ***reachability-distance*** of B w.r.t. C is defined as

   $$\text{reach-dist}_{\varepsilon, MinPts}(B, C) = max(\text{core-dist}_{\varepsilon, MinPts}(C), dist(C, B)) \quad .$$

---

1. Either (CF): execute BIRCH and extract the CFs.
   Or (SA): sample k objects from the database randomly and initialize k sufficient statistics.
   Classify the original objects to the closest sample object, computing sufficient statistics. Save classification information for use in the last step.
2. Compute Data Bubbles from the sufficient statistics.
3. Apply OPTICS to the Data Bubbles.
4. If (CF): classify the original objects to the closest Data Bubble.
5. Replace the Data Bubbles by the corresponding sets of original objects.

Figure 2.7: algorithm OPTICS-CF Bubbles and algorithm OPTICS-SA Bubbles

---

Figure 2.7 shows how OPTICS use the Data Bubbles concept to perform the clustering task. Step 1 is different for the two algorithms. For OPTICS-CF$_{Bubbles}$ [BKKS01] execute BIRCH, and extract the CFs from the leaf nodes of the CF-tree.

For OPTICS-SA Bubbles [BKKS01] draw a random sample of size and initialize a tuple *(n, L*S, *s*s) for each sampled object *s* with this object, i.e. *n=1*, *L*S=*s* and *ss* equals the square sum of s. Then, [BKKS01] read each object $o_i$ from the original database, classify $o_i$ to the sample object it is closest to, and update *(n, L*S, *s*s) for the corresponding sample point. [BKKS01] save the classification information by writing it to a file, as [BKKS01] can use it again in step 5. This is cheaper than to redo the classification.

In step 5, [BKKS01] reads each object $o_i$ and its classification information from the original database. Let $o_i$ be classified to $s_j$ and $B_j$ be the Data Bubble corresponding to $s_j$ .Now [BKKS01] set the position of $o_i$ to the position of $B_j$ .If $o_i$ is the first object classified to $s_j$ , [BKKS01] set th*e reachDist* of $o_i$ to the *reachDist* of $B_j$ ,otherwise [BKKS01] set the *reachDist* to *virtual-reachabilit*y(B). Then [BKKS01] write $o_i$ back to disc.

The time complexity of [BKKS01] is $O(k^2)$ which may be un acceptable for large *k*.

## 2.2 Hierarchal-based methods

The hierarchal-based methods put the data in a tree of clusters. The hierarchal-based clusters can be classified into agglomerative and divisive hierarchical clustering, depending on whether the decomposition is formed in a bottom-up or top-down manner.

In this section, we will introduce the well-known algorithms under this category.

**BIRCH**

BIRCH (Balanced Iterative Reducing Clustering and Using Hierarchies)[ZRL96] is an efficient hierarchical data clustering method for very large databases. BIRCH employees the following concepts.

1. Given $N$ $d$-dimensional data points in a cluster: $\{\vec{X}_i\}$ where $i = 1,...,N$, the **Clustering Feature** (CF) vector of the cluster is defined as a triple: $\mathbf{CF} = (N, \vec{LS}, SS)$, where $N$ is the number of the data points in the cluster, $\vec{LS}$ is the linear sum of the $N$ data points, i.e., $\sum_{i=1}^{N} \vec{X}_i$, and $SS$ is the square sum of the N data points, i.e., $\sum_{i=1}^{N} \vec{X}_i^{\,2}$.

2. A **CF** tree is a height-balanced tree with two parameters: branching factor $B$ and threshold T. Every non-leaf node consist of at most $B$ entries of the form [$CF_i$, $child_i$], where $i = 1,...,B$, "$child_i$" is a pointer to its $i$-th child node, and $CF_i$ is the CF of the sub-cluster represented by the this child. Every non-leaf node represents a cluster formed from all sub-clusters represented by its entries. A leaf node consists of at most $L$ entries each of the form [$CF_i$], where $i = 1,...,L$. Also each leaf node has two pointers "*prev*" and "*next*" that are used to connect all leaf nodes together in a chain for efficient scans. Also a leaf node represents a cluster made up of all the sub-clusters represented by its entries. But the diameter of every entry in a leaf node has to be less than the threshold $T$. Such a CF tree will be built dynamically as new data objects are inserted.

BIRCH algorithm consists of four phases. Fig. 2.8 presents the overview of BIRCH. The purpose of the first phase is to scan the data points and built an initial in-memory CF tree using the given amount of memory and recycling the space on the disk. The constructed CF tree reflects the clustering information of the database with crowded data points grouped as sub-clusters and the spare data points removed as outliers. The second phase is optional. It scans the leaf entries in the initial CF tree to rebuilt a smaller CF tree, while removing more outliers and grouping crowded sub-clusters into larger ones. The third phase uses a global or semi-global clustering algorithm to cluster all leaf entries. Actually the fact that the existing global or semi-global clustering methods applied in the third phase have different input size ranges within which they perform well in terms of both the speed and quality leads to the existence of the second phase as an optional one. The final phase uses the centriods of the clusters produced by the third phase as seeds to obtain a set of new clusters.

For a database *D* contains *n* objects, The complexity of BIRCH algorithm is O(*n*) as it only requires one scan.

Although the good clustering results, BIRCH doesn't perform well if the clusters are not spherical in shape. This because it uses the notion of diameter as a control parameter. Also Birch has problems recognizing clusters with different sizes.

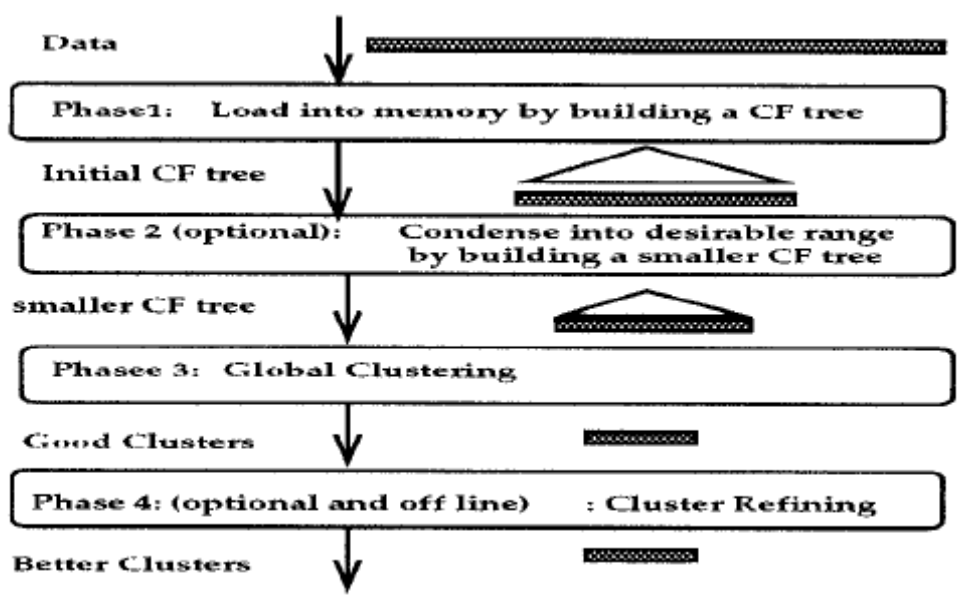

Figure 2.8: BIRCH Overview

**CURE**

CURE (Clustering Using Representatives)[GRS98] is an efficient clustering algorithm for large databases. CURE is a hierarchically agglomerative approach. CURE defines the problem of clustering as follows: given a data points in a d-dimensional metric space, partition the data points into *k* clusters such that the data points within a cluster are similar to each other than data points in different clusters.

CURE begins by considering each data point as a separate cluster. Then CURE iteratively merges the closest pair of clusters until the number of clusters becomes *k*. For each cluster, CURE stores *c*, which is a constant number, well-scattered representative point to be used in the computation of the distance between a pair of clusters. The determination of the *c* representative is done as follows. First, CURE chooses *c* well scattered points within the cluster. Then the chosen points are shrike toward the mean of the cluster by a fraction $\alpha$. To *c* well-scattered points in a cluster, CURE first chooses the point farthest from the mean as the first scattered point. Then CURE iteratively chooses a point from the cluster that is farthest from the previously chosen scattered points. The distance between two clusters is then the distance between the closest pair of representative points, one belonging to each of the two clusters.

Figure 2.9 shows CURE's approach to the clustering problem for large data sets.

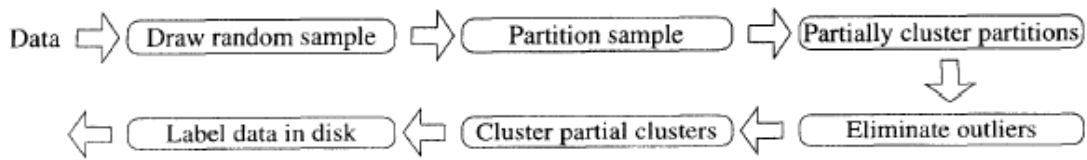

Figure 2.9: Overview of CURE approach for large databases.

The worst-case time complexity of CURE is $O(n^2 \log n)$. CURE has the following advantages. First, it can recognize arbitrary shaped clusters. Second, it handles outliers. Third, it has linear storage requirements. The other hand CURE has a major drawback that it fails to take into account special characteristics of individual clusters. For example consider the four sub-clusters of points in 2D sbown in Figure 2.10. The selection mechanism of CURE will prefer merging clusters (a) and (b) over merging clusters (c) and (d), since the minimum distances between the representative points of (a) and (b) will be smaller than those for clusters (c) and (d). But clusters (c) and (d) are better candidates for merging because the minimum distances between the boundary points of (c) and (d) are of the same order as the average of the minimum distances of any points within these clusters to other points. Hence, merging (c) and (d) will lead to a more homogeneous and natural cluster than merging (a) and (b).

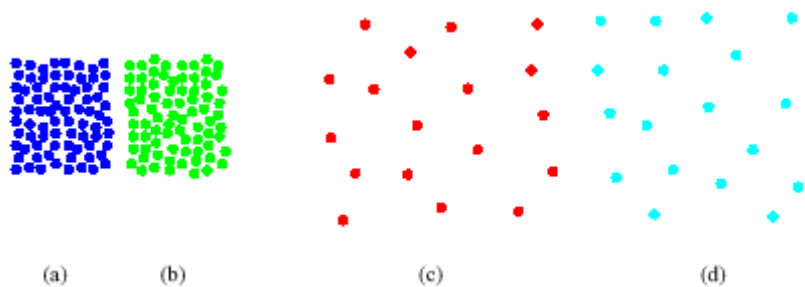

Figure 2.10: Example of clusters for mergeing choices

**CHAMELEON**

CHAMELEON [KHK99] is a hierarchal clustering algorithm Using Dynamic models. Figure 2.11 shows an overview of CHAMELEON framework.

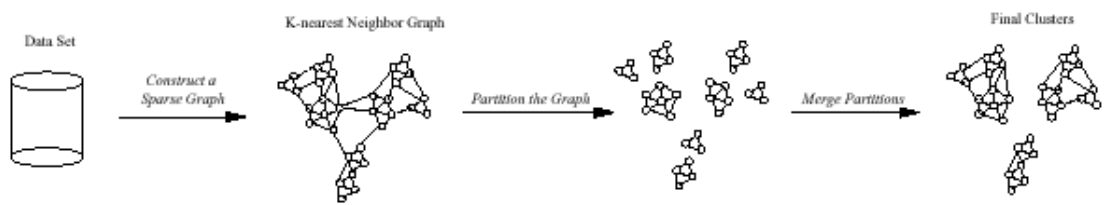

Figure 2.11: Overall framework CHAMELEON

CHAMELEON represents the database on the commonly used *k*-nearest neighbor graph. Each edge of the graph is weighted to indicate the degree of the similarity between the pair of the data items that are connected by that edge, i.e. an edge will weight more if the two data objects are similar toe ach other.

To determine the similarity between a pair of clusters $C_i, C_j$ CHAMELEON introduces the following new terminologies.

1. The ***relative inter-connectivity*** between a pair of clusters $C_i$ and $C_j$ is given by

$$RI(C_i, C_j) = \frac{|EC_{\{C_i, C_j\}}|}{\frac{|EC_{C_i}| + |EC_{C_j}|}{2}},$$

where $EC_{\{C_i, C_j\}}$ is the sum of the weight of edges that connect vertices in $C_i$ to vertices in $C_j$ and $EC_{C_i}$ is the weighted sum of the edges that partition the cluster $C_i$ into two roughly equal parts.

2. The *relative closeness* between a pair of clusters $C_i$ and $C_j$ is computed as by

$$RC(C_i, C_j) = \frac{\overline{S}_{EC_{\{C_i, C_j\}}}}{\frac{|C_i|}{|C_i| + |C_j|} \overline{S}_{EC_{C_i}} + \frac{|C_j|}{|C_i| + |C_j|} \overline{S}_{EC_{C_j}}},$$

where $\overline{S}_{EC_{C_i}}$ and $\overline{S}_{EC_{C_j}}$ are the average weights of the edges that belong in the min-cut bisector of clusters $C_i$ and $C_j$, respectively, and $\overline{S}_{EC_{\{C_i, C_j\}}}$ is the average weight of the edges that connect vertices in $C_i$ to vertices in $C_j$.

CAHMELEON consists of two phases. The purpose of the first phase is to cluster the data items into a large number of sub-clusters that contain a sufficient number of items. Using a graph-partitioning algorithm to partition the *k*-nearest neighbor graph of the data set into large number partitions such the edge-cut is minimized does this. The second phase uses an agglomerative hierarchical algorithm to discover the clusters in the sub-clusters produced by partitioning-based algorithm of the first phase. In this phase CHAMELEON starts with each of the sub-clusters produced in the first phase as a separate cluster. The algorithm then iteratively selects the most similar pair of clusters by looking at their relative inter-connectivity and their relative closeness to merge them. This done until a condition on the relative inter-connectivity and the relative closeness of the merged clusters is broken.

The total complexity of CHAMELEON 's two phases clustering algorithm is $O(nm + n \log n + m^2 \log m)$, where *n* is the number of the data items and *m* is the number of the sub-clusters produced by the first phase.

CHAMELEON has the following advantages. First, it can adapt to the internal characteristics of the clusters being merged. Second, it is more efficient than CURE in discovering arbitrary shaped clusters of varying density. But CHAMELEON can be time consuming for high dimensionality data items.

**ORCLUS**

ORCLUS [AY00](arbitrary Oriented projected CLUSter generation) is a hierarchical clustering algorithm that constructs clusters in arbitrary aligned subspaces of lower dimensionality. The subspaces are specific to the clusters themselves.

ORCLUS takes as input parameters the number of the clusters $k$ and the dimensionality $l$ of the subspace in which each cluster is reported. The algorithm outputs the following:

1. A $(k+1)$-way to partitions $(C_1,...,C_k,O)$ of data, such that the points in each partition element except the last form a cluster.
2. A possibly different orthogonal set $\varepsilon_i$ of vectors for each cluster $C_i$, $1 \leq i \leq k$, such that the points of $C_i$ cluster well in the subspace defined by those vectors.

The overall algorithm consists of a number of iterations, in each of which it applies a sequence of merging operations in order to reduce the number of current clusters by the factor $\alpha < 1$ and the dimensionality of the current clusters by the factor $\beta < 1$. $\alpha$ and $\beta$ are chosen such that the reduction from $k_o$ to $k$ clusters occur in the same number of iterations as the reduction from the $l_o = |D|$ (the dimensionality of the database) to $l$ dimensions.

The algorithm begins by picking a number $k_o > k$ of points from the database. These points are referred to as the seeds. At each stage of the algorithm, each seed $s_i$ is associated with the following:

1. Current cluster $C_i$: this is the set of the points from the database which are closest to seed $s_i$ in some subspace $\varepsilon_i$ associated with the cluster $C_i$.
2. Current Subspace $\varepsilon_i$: this is the subspace in which the points from $C_i$ cluster well.

For each of the initially chosen seeds $\varepsilon_i$ is initialized with the dimension set of the database. Then while the current number $k_c$ of seeds (clusters) is larger than $k$, the algorithm does the following. It calls the Assign procedure to construct the current clusters $C_1,...,C_{k_c}$ each of dimensionality $l_c$. It computes $k_{new} = max(k, k*\alpha)$ and $l_{new} = max(l, l_c*\beta)$. It calls the Merge procedure to reduce the number of the clusters from $k_c$ to $k_{new}$ clusters each of dimensionality $l_{new}$. Then it calls the FindVectors method procedure to find the set $\varepsilon_i$ of each of the new $k_{new}$ clusters. Then it puts $k_{new}$ as the current number of clusters and $l_{new}$ as the current dimensionality of clusters. The algorithm terminates merging process overall the iterations has reduced the number of clusters to $k$. The algorithm performs one final pass over the database in which it uses the Assign process in order to partition the database.

In each iteration the following three steps are applied:
1. Assign: The database is partitioned into $k_c$ current clusters by assigning each data point to its closest seed. The distance of the database point to a seed point $s_i$ is measured in the subspace $\varepsilon_i$.
2. FindVectors: In this procedure the algorithm find the subspace $\varepsilon_i$ of dimensionality $l_c$ for each current cluster $C_i$. This is done by computing the covariance matrix for the cluster $C_i$ and picking the $l_c$ orthonormal eigenvectors with the least eigenvalues.
3. Merge: During a given iteration the merge phase reduces the number of clusters from $k_c$ to $k_{new} = max(k, k_c*\alpha)$. The quantitative measure for the suitability of merging pair of seeds $[i,j]$ is calculated using a two step process. In the first step the eigenvectors corresponding to the smallest $l_{new}$ eigenvalues for the covariance matrix of $C_i \cup C_j$ is calculated. In the second step the mean square distance of then points in $C_i \cup C_j$ to its centroid is calculated in the. This calculation in the subspace founded in step1. So the pair with least square is merged first.

To make ORCULS scalable to very large databases, the ECF-vector is introduced. The ECF-vector contains $d^2 + d + 1$ entries. These entries are of three kinds:

1. There are $d^2$ entries corresponding to each pair of dimensions (*i*, *j*). For each pair of dimensions (*i*, *j*), the algorithm sums the products of the *i*th and the *j*th components for each point in the cluster.
2. There are $d$ entries corresponding to each dimension *i*. The algorithm sums the *i*th component for each point in the cluster.
3. The number of points in the cluster is stored also.

The total time required by the ORCULS algorithm is given by $O(k_o^3 + k_o \cdot N \cdot d + k_o^2 \cdot d^3)$. The overall space required of the algorithm is $O(k_o \cdot d^2)$.

The algorithm has the advantage of discovering the clusters in arbitrary oriented subspaces of lower dimensionality. Also it can prune off two many dimensions without at the same time incurring a substantial loss of information.

## 2.3 Partitioning methods

A partitioning algorithm divides *n* objects, which we want to cluster, into *k* partitions, where each partition represents a cluster and k is a given parameter. Such algorithm form the clusters to optimize an objective criterion, similarity function, such as distance.

In this section, we will introduce five well-known algorithms under this category.

**k-Means**

The *k*-Means[HK00] algorithm takes the input parameter, *k*, and partitions a set of *n* objects into *k* clusters so that the resulting intra clustering similarity is high but the intercluster similarity is low. Cluster similarity is measured in regard to the mean value of the objects in a cluster, which can be viewed as the cluster's center of gravity.

The *k*-means algorithm works as follows. First it randomly select *k* of the objects, each of witch initially represent a cluster mean. For each of the remaining objects, an object is assigned to the cluster to witch it is the most similar, based on the distances between the objects and the clusters centers. It then compute the new mean for each cluster. This process iterates until the criterion function converges. The sruared-error criterion is used, defined as

$$E = \sum_{i=1}^{k} \sum_{p \in c_i} |p - m_i|^2,$$

Where *E* is the sum for square-error for all objects in the database, *p* is the point in space representing a given object, and $m_i$ is the mean of the cluster $C_i$.

**k-Medoids**

The *k*-Means algorithm is sensitive to outliers since an object with extremely large value may substantially distort the distribution of the data.

Instead of taking the mean value of objects in a cluster as a reference point, the *k*-Mediods[HK00] algorithm use the mediod, which is the most centerly located object in a cluster. Thus the partitioning methods can still be performed based on the principle of minimizing the sum of the dissimilarity between each object and its corresponding reference point. This forms the bases of the *k*-Mediod method.

The basic strategy of the k-Mediods clustering algorithms is to find *k* clusters in n objects by first arbitrarily finding a representative object(the mediod) for each cluster. Each remaining object is clustered with mediod to which it is the most similar. The strategy the iteratively replaces one of the non-mediods as long as the quality of the resulting clustering is improved. This quality is estimated using accost function that measures the average dissimilarity between an object and the mediod of its cluster. To determine a non-mediod object, $o_{random}$, is a good replacement for a current object, $O_j$, the following four cases are examined for each of the non-mediod objects, $p$.

- Case 1: $p$ currently belongs to mediod $o_j$. if $o_j$ is replaced by $o_{random}$ as a mediod and $p$ is closest to one of the $o_i$, $i \neq j$ ,the $p$ is assigned to $o_i$.
- Case 2: : $p$ currently belongs to mediod $o_j$. if $o_j$ is replaced by $o_{random}$ as a mediod and $p$ is closest to $o_{random}$, then $p$ is assigned to $o_{random}$.
- Case 3: $p$ currently belongs to mediod $o_i$, $i \neq j$ . if $o_j$ is replaced by $o_{random}$ as a mediod and $p$ is still closest to $o_i$, then the assignment does not change.
- Case 4: $p$ currently belongs to mediod $o_i$, $i \neq j$ . if $o_j$ is replaced by $o_{random}$ as a mediod and $p$ is closest to $o_{random}$, then $p$ is assigned to $o_{random}$.

**CLARANS**

CLARANS (Clustering Large Applications based up on RANdomized Search) was introduced in [NG94] as the first clustering technique in spatial data mining problems. The algorithm takes as an input the number, $k$, of the desired clusters but such a parameter is often hard to determine in realistic applications. So a good clustering algorithm should minimize the input parameters. The algorithm first randomly selects $k$ points as the centers for the required clusters and assign each data point to its nearest center to form the required clusters. Then the algorithm tries to find better solutions. Better solution means a new set of centers that minimize the sum of the distances that each object has to cover to the center of its cluster. The computational complexity of CLARANS is $O(n^2)$, where $n$ is the number of the objects.

Although CLARANS generates good clustering results, there are several major problems with this algorithm. First, the quality of the results cannot be guaranteed when the number of points, N, is large since the randomized search is used in the algorithm to determine initial centers for the clusters and then to refine those centers.

Second, CLARANS takes as an input the number of the desired clusters and another integer, which determine the number of maximum tries to refine a center, but both numbers are generally unknowns in realistic applications. Third, CLARANS can't handle outliers. Forth, when data is updated, we need to run the algorithm from scratch. Finally, CLARANS assumes that all objects are stored in main memory. This clearly limits the size of the database to which CLARANS can be applied.

## 2.4 Grid-Based methods

The grid-based clustering algorithms summarize the data space into a finite number of cells that form a grid structure on which all of the operations for clustering are performed. Those methods always have fast processing time, which typically independent of the number of the data objects, but dependent on the number of the cells in each dimension in the summarized space.

In this section, we will introduce five well-known algorithms under this category.

**STING**

STING(STatistical INformation Grid )[WYM97] is a grid based clustering algorithm that is used to facilitate several kinds of spatial queries. The most commonly asked query is region query which is to select regions that satisfy certain conditions. An example of such a query is the following "Select the maximal regions that have at least 100 houses per unit area and at least 70% of the house prices are above $4OOK and with total area at least 100 units with 90% confidence". Another type of query selects regions and returns some function of the region, e.g., the range of some attributes within the region. An example of such a query is as follows " Select the range of age of houses in those maximal regions where there are at least 100 houses per unit area and at least 70% of the houses have price between $150K and $300K with area at least 100 units in California."

STING uses a hierarchical structure which constructed as follows. STING divides the spatial area into rectangle cells. Let the root of the hierarchy be at level 1; its children at level 2, etc. A cell in level *i* corresponds to the union of the areas of its children at level *i+1*. Each cell (except the leaves) has 4 children and each child corresponds to one quadrant of the parent cell. The root cell at level 1 corresponds to the whole spatial area. STING choose the size of the leaf level cells such that the average number of objects in each cell is in the range from several dozens to several thousands. The hierarchical structure is illustrated in Figure 2.12.

For each cell STING calculate the following parameters:
*n* - number of objects (points) in this cell
*m* - mean of all values in this cell
*s* - standard deviation of all values of the attribute in this cell
*min* - the minimum value of the attribute in this cell
*max* - the maximum value of the attribute in this cell
*distribution* - the type of distribution that the attribute value in this cell follows

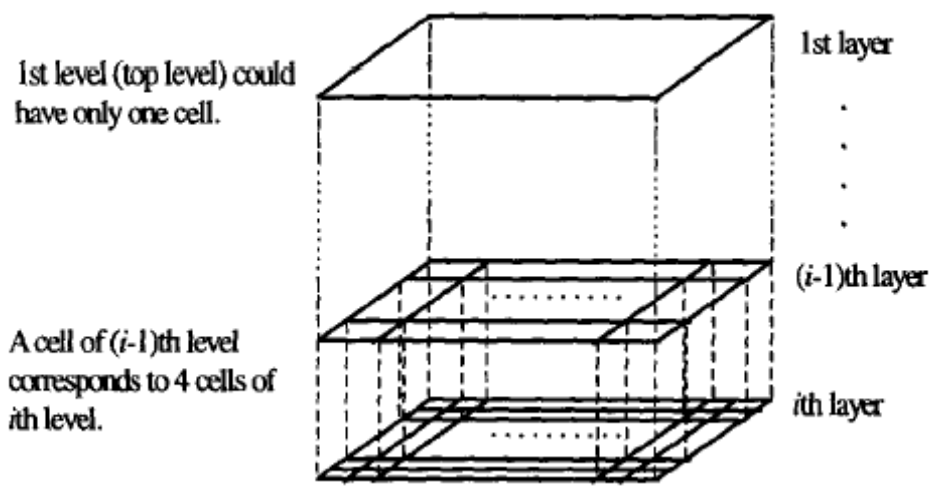

Figure 2.12: Hierarchical structure

After constructing the grid structure STING works as follows:
1. Determine a layer to begin with.
2. For each cell of this layer, STING calculates the confidence interval (or estimated range) of probability that this cell is relevant to the query.

3. From the interval calculated above, STING labels the cell as relevant or not relevant.
4. If this layer is the bottom layer, go to Step 6; otherwise, go to Step 5.
5. STING goes down the hierarchy structure by one level. Go to Step 2 for those cells that form the relevant cells of the higher level layer.
6. If the specification of the query is met, go to Step 8; otherwise, go to Step 7.
7. Retrieve those data fall into the relevant cells and do further processing. Return the result that meet the requirement of the query. Go to Step 9.
8. Find the regions of relevant cells. Return those regions that meet the requirement of the query. Go to Step 9.
9. Stop.

The total complexity of STING is *O(K)*. Where K is the number of cells at bottom layer.

Although STING produces a good clustering results, it has the following drawbacks. First STING quality depends on the granularity of the lower level cells. Second, STING does not consider the regions formed by children of different parents. As a result the diagonal bounders of a cluster can't appear.

**WaveCluster**

WaveCluster [SCZ89] is a multi-resolution clustering approach for very large databases. The basic idea beyond this approach is as follows. The multi-dimensional spatial data objects can be represented in an *n*-dimensional feature space. A feature vector represents the numerical attributes of a spatial object where each element of the vector corresponds to a numerical attribute. Each object with *n* numerical attributes corresponds to one point in the feature space. WaveCluster propose to look at the feature space from a signal processing prospective. So the high frequency parts of the signal correspond to the regions of the feature space where there is a rapid change in the distribution of the objects, that is the boundary of the clusters. The low frequency parts of the *n*-dimensional signal, which have high amplitude, correspond to the areas of the feature space where the objects are concentrated, i.e. the clusters themselves.

To find the different frequency sub-bands of a signal, WaveCluster employees a wavelet transform. Wavelet transform is a type of signal representation that can give the frequency content of the signal at a particular instance of time. By applying wavelet transform multiple times, WaveCluster can detect the clusters at at different levels of accuracy. For example, Figure 2.13 shows the wavelet representation of a database at three scales from fine to coarse.

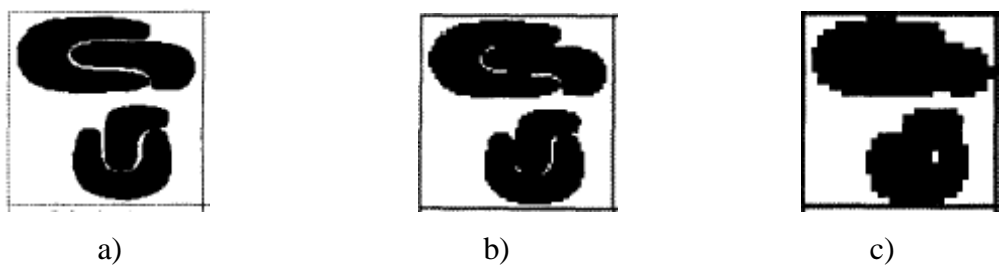

a) b) c)

Figure 2.13: Multi-resolution wavelet representation of a database at a) scale 1; b) scale2; c) scale 3.

The WaveCluster algorithm works as follows. It quantizes feature space and assigns objects to the units. Then, it applies wavelet transform on the feature space. It finds the connected components in the sub-bands of transformed feature space at different levels. Then, it assigns the labels to the units. The algorithm maps the units in the transported feature space to the units in the original space. Finally, it maps the objects to the clusters.

The required time for WaveCluster algorithm is linear in terms of the number of the objects in the database.

WaveCluster has the following advantages. First, it is order insensitive with respect to the input objects. Second, it can detect arbitrary shape clusters. Third, it removes the outliers from the original space after the transformation. Finally, by using wavelet transform, Clusters in the data automatically stand out and clear regions around them.

**CLIQUE**

CLIQUE(CLustering in QUEst)[AGGR98] is a grid-based clustering algorithm. CLIQUE, has been designed to find clusters embedded in subspaces of high dimensional data without requiring the user to guess subspaces that might have interesting clusters. CLIQUE generates cluster descriptions in the form of DNF expressions that are minimized for ease of comprehension. It is insensitive to the order of input records and does not presume some canonical data distribution.

CLIQUE, consists of the following steps:
1. Identification of subspaces that contain clusters.
2. Identification of clusters.
3. Generation of minimal description for the clusters.

The first step is done as follows. It first determines 1-dimensional dense units by taking a pass over the data. Having determined *(k-1)*-dimensional dense units, the candidate *k*-dimensional units are determined using the candidate generation procedure given below. A pass over the data is made to find those candidate units that are dense. CLIQUE terminates when no more candidates are generated.

The candidate generation procedure takes as an argument $D_{k-1}$, the set of all *(k-1)*-dimensional dense units. returns a superset of the set of all k-dimensional dense units. Assume that the relation < represents lexicographic ordering on attributes. First, CLIQUE self-join $D_{k-1}$, the join condition being that units share the first *k-2* dimensions. CLIQUE then discard those dense units from $C_k$ which have a projection in *(k-1)*-dimensions that is not included in $C_{k-1}$.

In the second step CLIQUE uses a depth-first search to find the connected components of the dense units in the obtained sub-space.

The input to the third step consists of disjoint sets of connected *k*-dimensional units in the same subspace. Each such set is a cluster and the goal is to generate a concise description for it. To generate a minimal description of each cluster, CLIQUE would want to cover all the units comprising the cluster with the minimum number of regions such that all regions contain only connected units. For a cluster *C* in a *k*-dimensional subspace *S*, a set *R* of regions in the same subspace *S* is a cover of *C* if every region $R' \in R$ is contained in C, and each unit in C is contained in at least one of the regions in *R*.

CLIQUE is linear in the number of the database objects. CLIQUE is scalable, end-user comprehended of the results. CLIQUE also doesn't presume any canonical data distribution, and insensitivity to the order of input records.

**STING+**

STING+[WYM99] supports spatial data mining triggers monitoring both spatial regions that satisfy some condition and attribute values of objects within some spatial regions. The user can specify two kinds of conditions. One condition is an *absolute* condition, i.e., the condition is satisfied when a certain state is reached. The other type of condition is a *relative* condition, i.e., the condition is satisfied when a certain degree of change has been detected.

Therefore, four categories of triggers are supported by STING+ .
1. **region-trigge**r: *absolute* condition on certain regions,
2. **attribute-trigge**r: *absolute* condition on certain attributes,
3. **region**-$\delta$-**trigge**r: *relative* condition on certain regions,
4. **attribute**-$\delta$-**trigge**r: *relative* condition on certain attributes.

STING+ employees a hierarchical structure that similar to the structure used by STING. Also there are sex types of variables known as sub-triggers. These sub-triggers are added to different cells according to the user-defined trigger being handled.

The sex types of sub-triggers are the following.
1. *Insertion-sub-triggers* and *deletion-sub-triggers* are sub-trigger types (referred to as *density-sub-trigger*s) used to monitor density changes of a cell.
2. Another pair of sub-triggers (referred to as *attribute-sub-trigger*s) are *inside-sub-trigger* and *outside-sub-trigge*r. They monitor whether an aggregate (such as MIN, MAX, AVERAGE, etc.) or a certain percentage (e.g., 80%) of attribute values enters or leaves a range $[r_l, r_u]$ respectively.

3. Moreover, STING+ allows two additional types of sub-triggers *expand-sub-trigger* and *shrink-sub-trigger* to be set *only* on leaf level cells to track expansion and shrinking of region(s) $Q$ from a given time, $t_1$, respectively.

Then, there are two sets of composite events that STING+ considers: (1) the set, $E(s)$, of composite events that can cause $C_T$ to become true, where $C_T$ is the trigger condition specified in *T(trigger)*; (2) the set of composite events that can cause a change to $E(s)$, call it $F(s)$. $(F(s)$ is closed, i.e., only events in $F(s)$ can update $F(s)$. If a composite event in $E(s)$ occurs, we need to re-evaluate $C_T$; whereas a composite event in $F(s)$ happens, we have to update $E(s)$.

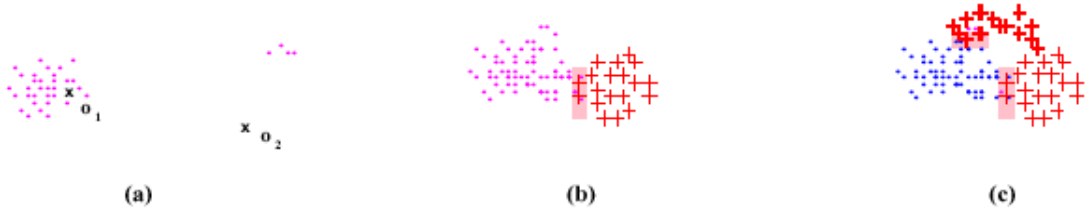

Figure 2.14: Effect of events Figure

To evaluate $C_T$, STING+ employs a "step-by-step" strategy based on the observation that a trigger condition is a conjunction of predicates $P_1 \cap P_2 \cap ... \cap P_n$ and can not become true if any one predicate is false. In particular, the trigger evaluation process in STING+ employs the order {location, density condition, attribute condition,...} and is therefore divided into phases, one for each predicate.

In many applications, clusters formed by objects with some specified attribute values are the main features of interest. Some examples are the following.

1. *Military Deployment*.
2. *Situation Awareness and Emergency Response*.

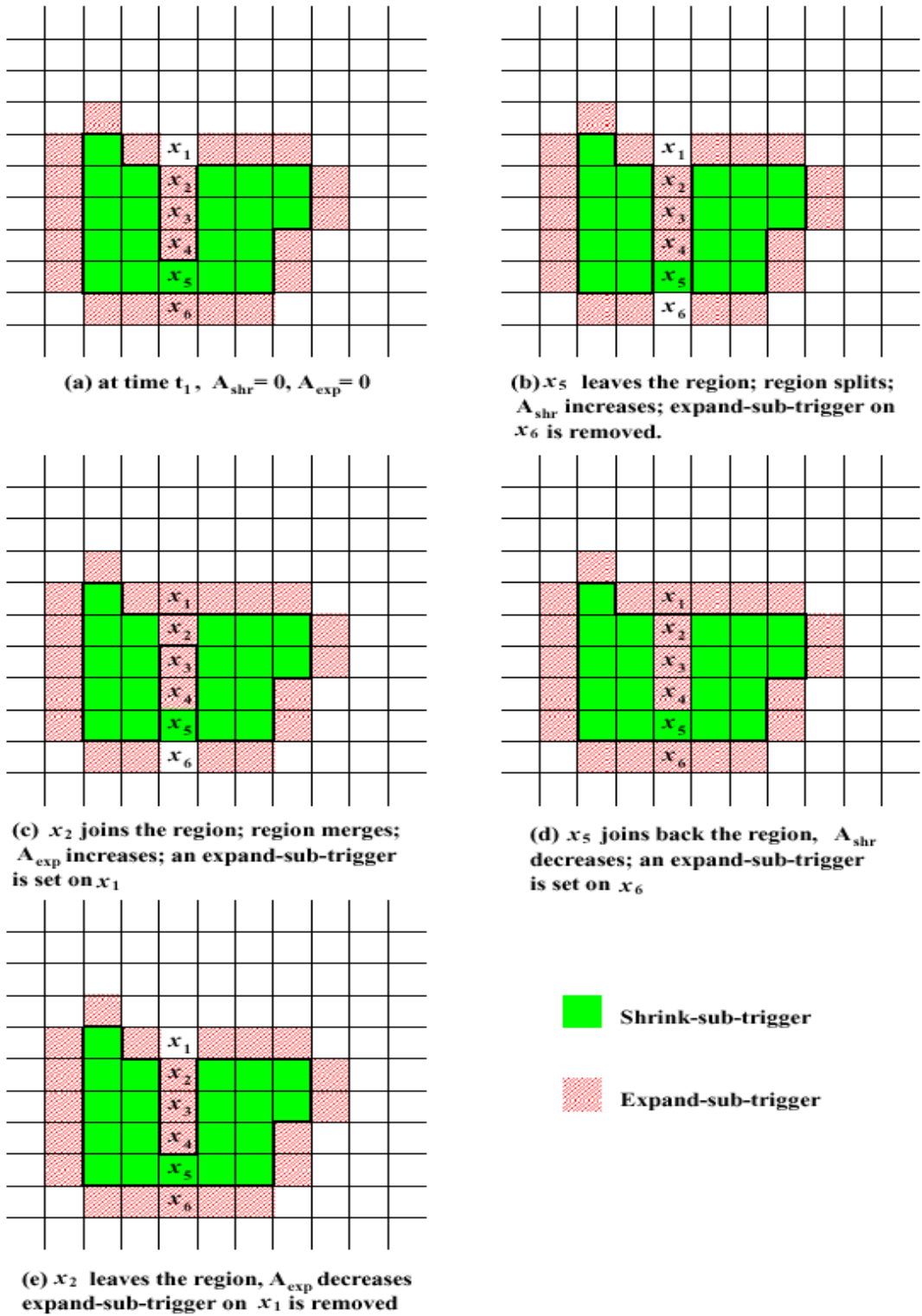

Figure 2.15: shrank and expand triggers

## 2.5 Constrained-based methods

Many clustering algorithm have been conducted. Most of these algorithms does not allow users to specify real life constraints such as physical obstacles. Additional research is needed on providing the users the ability to incorporate real life constraints into the clustering algorithm.

In this section, we introduce two algorithms which are of such nature.

**Constrained-Based Clustering in Large Databases**

The constrained clustering problem is defined in Constrained-Based Clustering in Large Databases [THLN01] as follows.

**Constrained Clustering (CC):** Given a data set *D* with *n* objects, a distance function $df : D \times D \rightarrow \Re$, a positive integer *k*, and a set of constraints *C*, find a *k*-clustering, i.e. a partition of *D* into *k* disjoint clusters $(Cl_1,...,Cl_k)$ such that $DISP = (\sum_{i=1}^{k} disp(Cl_i, rep_i))$ is minimized, and each cluster $Cl_i$ satisfies *C*, denoted as $Cl_i \models C$. Where $rep_i$ is the representative of the cluster $Cl_i$ and is chosen such that the dispersion of the cluster $Cl_i$, $disp(Cl_i, rep_i)$, which equals $\sum_{p \in Cl_i} df(p, rep_i)$ is minimized.

Constrained-Based Clustering in Large Databases [THLN01] develops a scalable constrained clustering algorithm to solve the constrained clustering problem. The algorithm mainly focuses on a type of constraints called **existential constraints** that is defined as follows.

**Existential Constraints:** Let $W \subset D$ be any subset of objects. We call them **pivot** objects. Let *c* be a positive integer. An existential constraint on a cluster *Cl* is a constraint of the form: $count(\{O_i | O_i \in Cl, O_i \in W\}) \geq c$.

The algorithm finds an initial solution satisfying user-defined constraints and then tries to refine the solution by doing confined object movements under constraints. The algorithm is an iterative one that consists of two phases: *pivot movement* and *deadlock resolution*. Although the refinement process is done during the two phases, phase two is mainly designed to resolve a deadlock cycle. Briefly, a sequence of clusters $<Cl_1,...,Cl_k,Cl_1>$ is said to be in a deadlock cycle of length $k$ if (a) the movement of any object from any cluster will violet the constraints; and (b) there exist an object in the cluster $Cl_i$ with $rep_{i+1}$ as its nearest cluster representative, $1 \leq i \leq k-1$, and one object in $Cl_k$ with $rep_1$ as its nearest cluster representative. In [THLN01], it is proven that finding the optimal solution for both phases is NP-hard problem so several heuristics are used in the algorithm.

Constrained-Based Clustering in Large Databases presents a novel concept called *micro-clustering sharing* to scale up the algorithm. Given a micro-cluster (a compressed entity of data objects) with $n$ non-pivot objects and $m$ pivot objects, the $n$ non-pivot objects will always be assigned to the nearest cluster, while $m$ pivot objects can be shared among the multiple clusters.

Constrained-Based Clustering in Large Databases also discusses other types of constraints: *universal, existential-like, averaging, and summation*. Universal constraints are constraints in which a specific condition must be satisfied by every object in a cluster. This can be reduced to the unconstrained clustering problem by running the clustering algorithm on only those objects that satisfy the constraints. Existential-like constraints are similar to existential constraints in nature and can be handled by an algorithm that was designed to handle existential constraints with simple modification. For example, $count(Cl_i) \leq c$ is an existential-like constraint. Averaging and summation constraints are constraints, which involves summation or averaging of some numerical attributes of the data objects. Even finding an initial solution for averaging and summation constrain is an NP-hard problem.

**COD-CLARANS**

COD-CLARANS[THH01](Custering with Obstructed Distance based on CLARANS) is the first clustering algorithm that solve a problem which is known as the problem of clustering with obstacles entities(COE).

The problem of clustering with obstacles entities(COE) is to partition a set of n points $P = \{p_1,...,p_n\}$ into $k$ clusters $Cl_1,...,Cl_k$ such that the following square error function $E$ is minimized.

$$E = \sum_{i=0}^{k} \sum_{p \in Cl_i} (d'(p,c_i))^2$$

Given that the space contains *m* non-intersect obstacles $\{o_1,...,o_m\}$ in a two-dimensions region, *R,* with each obstacle represented by a simple polygon.

The COD-CLARANS algorithm contains two phases. The first phase breaks the database down to several sub-databases and summarizes them individually. For each clustering step a micro-clustering step is performed. A micro-cluster is a compressed representation of a group of points which are close together that they are likely to belong the same cluster. Information about the micro clusters are stored together with the representative points. A BSP tree and a visibility graph are also constructed during this phase. Spatial join indexes such as VV index, MV index, or MM index can be calculated also during this phase for computing the obstructed distance efficiently.

The second phase is the clustering stage. The algorithm first randomly selects *k* points as the centers of the clusters and then ties to find better solution.

The figure above shows the overall structure of COD-CLARANS. The algorithm consists of three main parts the main algorithm, the computation of the square-error *E* , and a pruning function *E'*. The pruning function *E'* has two purposes. First it can help to prune search and avoid the computation of *E*. Second in the event when the computation of *E* can't be avoided, the pruning function can provide focusing information to make the computation of E more efficient.

In order to discuss the computational complexity of COD-CLARANS, we suppose that *D* is a database with *n* objects, *m* is the number of the micro-clusters to be clustered and *G(V,E)* is the visibility graph of the polygon obstacles. The expected running time of the first phase is $O(N\ log|V|)+O(|V|^2)$. The expected running time of second phase is $O(N |V|)+O(m|V|^2)+ O(m^2|V|)$. So the total complexity of the COD-CLARANS algorithm is

$$O(COD\text{-}CLARANS)=O(N |V|)+O(m|V|^2)+ O(m^2|V|)$$

Although COD-CLARANS generates good clustering results, there are several major problems with this algorithm. First, the quality of the results cannot be guaranteed when the number of points, N, is large since the randomized search is used in the algorithm to determine initial centers for the clusters and then to refine those centers. Second, COD-CLARANS takes as an input the number of the desired clusters and another integer, which determine the number of maximum tries to refine a center, but both numbers are generally unknowns in realistic applications. Third, COD-CLARANS can't handle outliers. Forth, when data is updated, we need to run the algorithm from scratch.

```
Input: A set of n objects, k and clustering parameters, maxtry.
Output: A partition of the n objects into k clusters with cluster
centers, c_1, ..., c_k.
Method:

1. Function COD-CLARANS()
2. { randomly select k objects to be current;
3.   compute square-error function E;
4.   let currentE = E;
5.   do
6.   { found_new = FALSE;
7.     randomly reorder current into {c_1,...,c_k};
8.     for (j=1 ; j≤k ; j++)
9.     { let remain = current − c_j ;
         /* remain contain the remaining center */
10.      compute obstructed distance of objects to nearest
         center in remain;
11.      for (try=0; try < max_try; try++)
12.      { replace c_j with a randomly selected object c_random ;
13.        compute estimated square-error function E';
14.        if (E' > currentE)
15.          continue; /* Not a good solution */
16.        compute square-error function E;
17.        if (E < currentE) /* Is the new solution better ? */
18.        { found_new = TRUE; /* Found a better solution */
19.          current = {c_1,..., c_random, ...c_k}
           /* replace c_j with c_random */
20.        currentE = E;
21.      }
22.    }
23.    if (found_new)
24.      break; /* Reorder cluster centers again */
25.  }
26. } while (found_new)
28. output current ;
27.}
```

Figure 2.16: The COD-CLARANS algorithm

# Chapter 3

# Clustering Large Spatial Databases

The requirements of clustering algorithms which are raised by applications of large spatial databases are the following: minimal requirements of domain knowledge to determine the input parameters, efficient discovery of clusters with arbitrary shapes , and acceptable efficiency in handling large databases. Most algorithms that solve the problem of spatial clustering, however, do not present solutions that satisfy all requirements. As mentioned in the Introduction, one of the objectives of this thesis is to propose a clustering algorithm for large spatial databases that satisfy all the aforementioned requirements.

In this chapter, we introduce a grid-based clustering algorithm. We first introduce some notions that are used in the algorithm. In section 3.2, we present the details of the algorithm. In section 3.3, we introduce an illustrative example. Section 3.4 presents the computational complexity of the algorithm. An experimental evaluation of the proposed algorithm using synthetic data is presented in Section 3.5. Finally, Section 3.6 concludes the chapter.

## 3.1 Basic Definitions

In the following, we formalize the necessary notions for our proposed algorithm.

**Definition 1:** (a grid structure of a spatial area) Given a spatial area *S* a ***grid structure*** of *S* is a structure resulting from dividing *S* into rectangular cells of equal areas. These cells are obtained by dividing the dimensions of the spatial area into the same number of the segments that have equal lengths. Figure 3.1 shows an example of such grid structure.

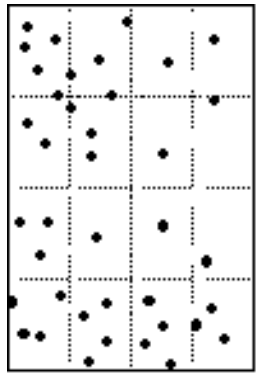

Figure 3.1: the grid structure for a spatial database

**Definition 2:** (dense cell) A cell in the grid structure is said to be a ***dense cell*** if the number of points in that cell is greater than or equal $\frac{N}{m} * h$. Otherwise the cell is said to be ***non-dense.*** Where *S is* a grid structure of a spatial area, *N* is the number of points in the spatial area, *m* is the number of the cells in the grid structure, *h* is a proportion.

**Definition3:** (neighbor cell) Given a grid structure of a spatial area, two cells are said to be ***a neighboring cells***, if the two cells has a common boundary (either in a line or a point).

**Definition 4:** (connected cells) Given a grid structure of a spatial area, two cells, *p* and *q*, are said to ***connected*** if there exists a chain of cells $c_1,...,c_n$, where *n* is a natural number, such that $c_1 = p, c_n = q$ and $c_i$ is a neighboring cell to $c_{i+1}, 1 \leq i \leq n-1$.

**Definition 5:** (connected region) Given a grid structure of a spatial area, *S*, a ***region*** in *S* is a set of cells. A region is said to be ***connected,*** if any pair of cells in the region are connected. A region of *dense* cells is said to be ***maximal*** if any *dense* cell that is a neighbor to a cell in that region is also in the region.

**Definition 6:** (cluster) Let *S* be a grid structure of a spatial area, *N* is the number of points in the spatial area, *m* is the number of the cells in the grid structure, and *h* is a proportion, a ***cluster*** in the spatial area *S* is a connected region of cells such:

1. the number of points in the cluster is greater than or equal to $\left(\dfrac{N}{m} * h\right) * t$, where $t$ is the number of cells in the cluster.

2. Addition of a cell that is a neighbor to some cell in the cluster violates the first condition.

**Definition 7:** (outliers) Let $C_1,...,C_k$ be the clusters in a spatial area $S$ with respect to the parameters $m$, the number of the cells in the grid structure, $h$, a proportion, then we define the ***outliers*** as the set of data points in the spatial area that do not belong to any cluster $C_i$.

## 3.2  Algorithm *SCLD*

In this section, we describe the proposed algorithm, *SCLD*. The *SCLD* algorithm consists of two phases.

### 3.2.1  Phase 1

In this phase, the algorithm first divides the spatial area into $m$, which is an input, rectangular cells of equal areas by dividing each of the dimensions of the spatial area into the same number, $w = \sqrt{m}$, of equal pieces. Then, the algorithm labels each cell as *dense* or *non-dense* (according to the number of points in that cell and an input threshold). The algorithm finds all maximal, connected, regions of *dense* cells. Each region is a cluster. Then, the algorithm finds a center for each of the obtained clusters. To find a center for a cluster the algorithm determines the arithmetic mean of all the points in the cluster.

### 3.2.2 Phase 2

In this phase, for each *non-empty*, *non-dense* cell, the algorithm assigns the cell to its neighboring cluster that satisfy the following conditions. First, the addition of the cell to the cluster dose not violate the first condition in the definition of the cluster (see Definition 6). That is the number of the points in cluster after adding this cell, is greater than or equal to $\left(\frac{N}{m}*h\right)*(t+1)$. Second, if there are more than one cluster that satisfy the first condition, the cell is added to the one whose center in the nearest (using the Euclidean distance) to the mean point of the cell. Finally the algorithm re-computes a center for each extended cluster and outputs the obtained centers as the centers of the clusters in the spatial area.

*Algorithm 3.1 SCLD.*

*Input*:

1. A set of *N* objects in a spatial area *S*.
2. A square number, *m*, which represents the number of cells in the spatial area such that $m<<N$.
3. A percentage, *h*, used to determine the dense cells according to definition 2.

*Output*: Clusters with their centers.

*Method*:

1    $w = \sqrt{m}$;

2   Divide the spatial area into *m* rectangular cells by dividing each of dimension of the spatial area into *w* equal segments;

3   For(*i*=0; *i*<*m*; *i*++)

      {

    determine for the cell, $c_i$, parameters:

-   $n_{c_i}$ : the number of the objects in that cell.
-   $m_{c_i}$ : the mean object of all objects in that cell.

    }

4   $d = round(\frac{N}{m} * h)$;

5   For(*i*=0; *i*<*m*; *i*++)

    {

   if ($n_{c_i} \geq d$) then

       $c_i$ is labeled as *dense;*

   else

       $c_i$ is labeled as *non-dense;*

    }

6   *j*=0;

7   For(*i*=0; *i*<*m*; *i*++)

    {

   if ($c_i$ is dense) then,

     {

    if ($c_i$ is not processed yet) then,

       {

         - Construct a new cluster, $r_j$, and mark the cell $c_i$ as an element of $r_j$;

         - Put the *dense* neighboring cells of $c_i$ in a list, *Q;*

        - While ($Q$ is not empty)

          {

          - Take the first element, $c'$, from $Q$, mark $c'$ as an element of the cluster $r_j$ and add to $Q$ the *dense* neighboring cells of $c'$, that have not been processed yet;

          }

        - $j$++;

      }

    }

8    For($i=0$; $i<j$; $i$++)

        Compute the mean object of the cluster $r_j$;

9    For($i=0$; $i<j$; $i$++)

      {

      - List all neighboring non-empty cells of $r_j$ in a list, $QN$;

      - Sort, $QN$ in a descending order, according to the number of objects in the cells;

      }

10  While ($QN$ isn't empty)

      {

      - Take the first element, $e$, of $QN$;

      - Find all neighboring clusters of $e$;

      - Determine which of those clusters can be extended to contain $e$ without becoming non-dense;

      - Assign $e$ to the nearest cluster obtained in previous step;

      }

11  Re-compute a center for each extended cluster;

12  Output the clusters with their centers;

## 3.3 Example

In this section, we describe our algorithm by an example. Suppose that we have a spatial area containing 5,000 objects. Figure 3.2 shows a sketch of the spatial area.

We are asked to find the clusters in that spatial area given that $m = 36$ and $h = 0.9$.

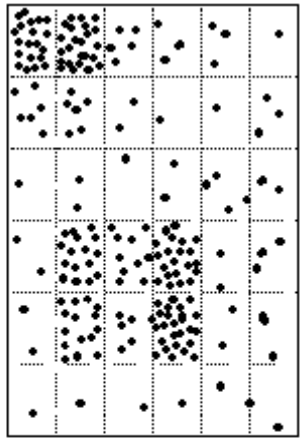

Figure 3.2: a skech of a spatial area with objects represented by points.

The algorithm works as follow.
1. $w = \sqrt{m} = 6$.
2. The spatial area is divided into 36 equal cells by dividing each of the dimensions into 6 equal segments. So we obtained the dotted grid structure in Figure 3.2. The cells in Figure 3.2 are numbered from left to right and from top to down.

3. For each cell, *c*, the algorithm computes the two parameters $n_c$, $m_c$. Table 3.1 shows values for the number of points for some cells in the grid. We assume that the remaining points in the spatial area are distributed among the remaining cells such that no cell contains more than 100 points.

4. $d = round(\frac{N}{m} * h) = 125$.

5. Each cell, c, with $n_c \geq d$ is labeled as *dense*. So, cells 1, 2, 14, 16, 20 and 22 are the only *dense* cells.

| Cell number | Number Of points |
|---|---|
| 1 | 400 |
| 2 | 400 |
| 20 | 125 |
| 21 | 120 |
| 22 | 600 |
| 26 | 125 |
| 27 | 110 |
| 28 | 600 |

Table 3.1: the $n_c$ 's parameters for some cells

6. The maximal, connected regions (clusters) of *dense* cells are determined by a breadth-first search. Figure 3.3 illustrates such clusters in the given spatial area with each cluster in a different color.

7. For each cluster obtained in step 6, the algorithm finds the arithmetic mean of the points in the cluster as its center. These centers are shown in Figure 3.3.

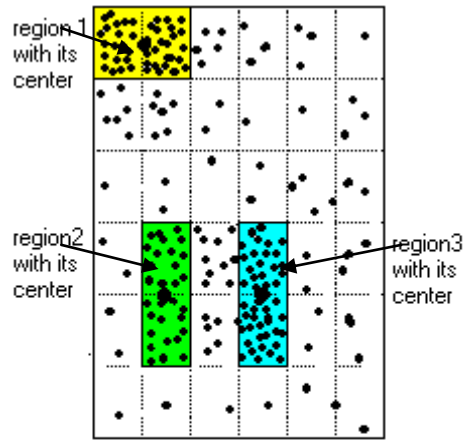

Figure 3.3: The regions founded after step 6 with their centers

8. In this step, cells 3, 7, 8, 9, 13, 14, 15, 16, 17, 19, 21, 23, 25, 27, 29, 30, 31, 32, 33, 34 and 35 that are neighbors to the obtained clusters are put in a list *QN* in a descending order according to the number of the points in each of them. So cell 21 is the first element of *QN* followed by cell 27.

9. The algorithm proceed as follows:

    9.1 Cell 21 is eliminated from *QN* because this cell at the top of the list *QN* .

    9.2 Regions 2, 3 are determined as neighboring clusters to cell 21.

    9.3 Region 3 is determined. In this step region 2 is eliminated because the addition of cell 21 to region 2 violate the first condition in the definition of the cluster (Definition 6 ).

    9.4 Cell 21 is assigned to region 3. Steps 9.1 to 9.4 are repeated to each element of *QN*.

10. Figure 3 shows the extended clusters with their new centers.

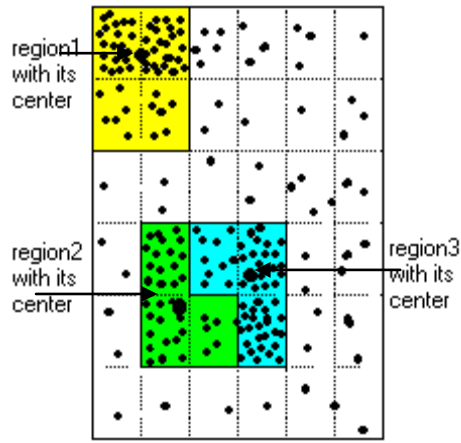

Figure 3.4: The resulted regions with their centers

## 3.4  Complexity of the *SCLD* Algorithm

We analysis the complexity of the *SCLD* in this section. The following notations are defined for this discussion:

*N*: the number of data objects in database *D*.

*m*: the number of cells in the spatial area, $m<<N$.

Steps 1, 2, 4, 7, and 10 are simple calculations that take a constant time. In step 3, we need to scan the database only once so step 3 takes *O(N)* time. Step 5 takes *O(m),* because in this step we need to scan the cells in the grid structure only once. The expected running time of step 6 is also *O(m)*. So, the running time for steps 5 and 6 is *O(m)+O(m)= O(m)* time. The expected running time of step 8 is *mlog( m )*, because we need to order some of the cells in the grid structure according to their corresponding number of the data points. Step 9 takes time of *O(m),* because in this step we scan some of the cells in the grid structure.

The total complexity of *SCLD* is the sum of the running time of all the steps:

$$O(SCLD) = O(N) + O(m) + m\log(m) + O(m).$$
$$= O(N) + m\log(m) \text{ (since } m<<N\text{)}$$

## 3.5 Performance Study

In this section, we look at the performance of the *SCLD* algorithm by performing experiments on a PC with a Pentium 900 Mhz processor and a Western digital 750 rpm hard disk. We use two synthetic databases, DS1 and DS2. These databases are shown in Figure 3.5. In DS1, there are five clusters of different shapes and sizes. In DS2, there are four circle-shaped clusters of significantly differing sizes. DS1 contains additional noise. The number of data points in DS1 and DS2 are 42000 and 40000, respectively.

The experiments proceeded as follows. First, we assess the efficiency and effectiveness of the algorithm, by running it on DS1 for different values of the parameter *m* which represents the number of the formed cells in the spatial area.

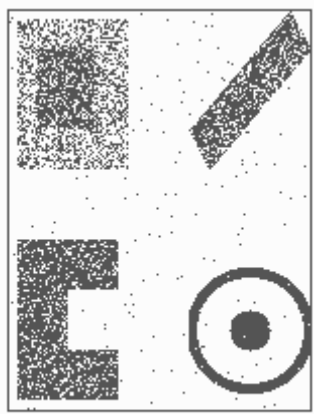
(a) DS1.

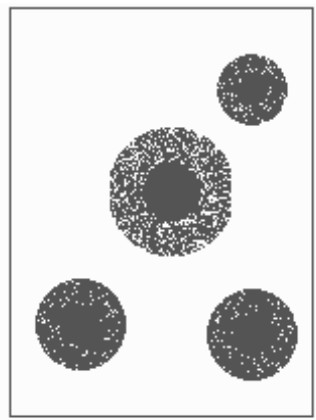
(b) DS2.

Figure 3.5: Spatial databases.

Second, we compare *SCLD* with the performance of *CLARANS* [NG94] which is one of the most robust clustering algorithms.

**Varying *m***

In this experiment, apply the algorithm on the database DS1 with different values of *m*. There are two reasons in testing the algorithm with different *m* values. First, because more accuracy will be granted when *m* is increased, we like to investigate how the quality of the clusters is affected by the choice of *m*. Second, to insure the impact of the choosing of *m* on the running time of the algorithm. To show the results of the clustering process, we visualize each cluster by a different color. Figure 3.6 shows the clustering results for running *SCLD* on DS1 for different values for *m*. From the figure, we note that as we increase *m* the quality of the clustering process is increased by eliminating more noise points.

Table 3.2 shows the running times in seconds, of *SCLD* on Ds1 for different values of *m*.

From the table, we note that the time does not increase dramatically as *m* increases.

| $M$ | Running time in seconds |
|---|---|
| 484 | 0.49 |
| 576 | 0.50 |
| 676 | 0.55 |
| 784 | 0.56 |
| 900 | 0.57 |
| 1225 | 0.59 |

**Table 3.2: The running time for different values of *m***

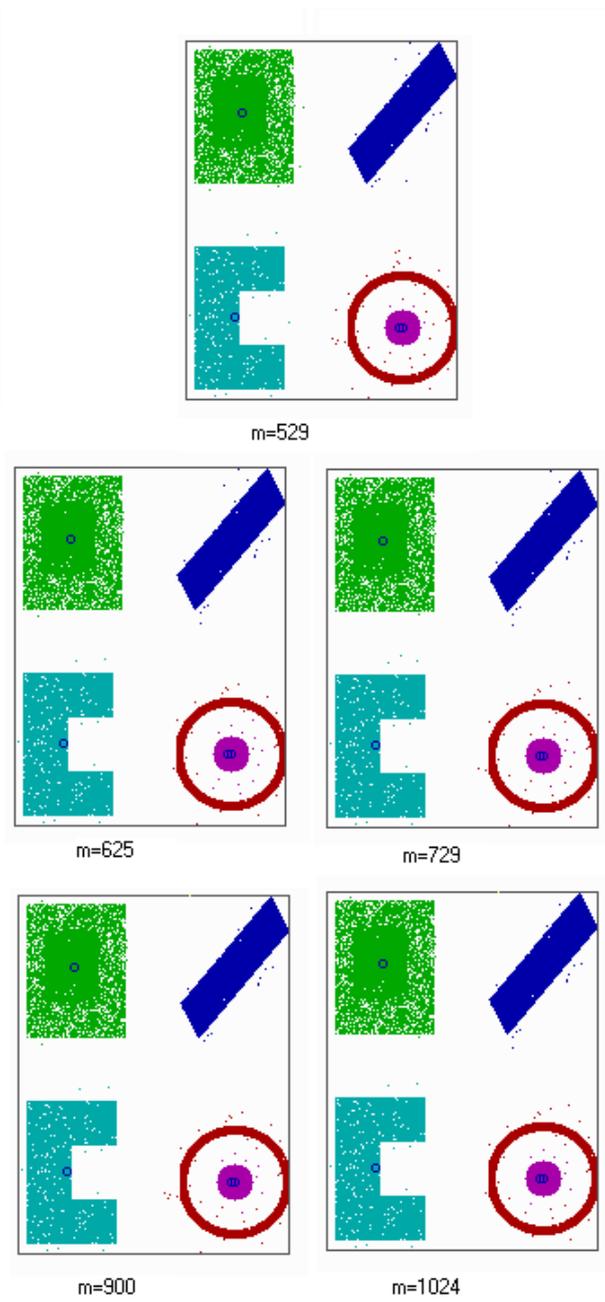

Figure 3.6: The cluster results by SCLD on DS1 for different values on m

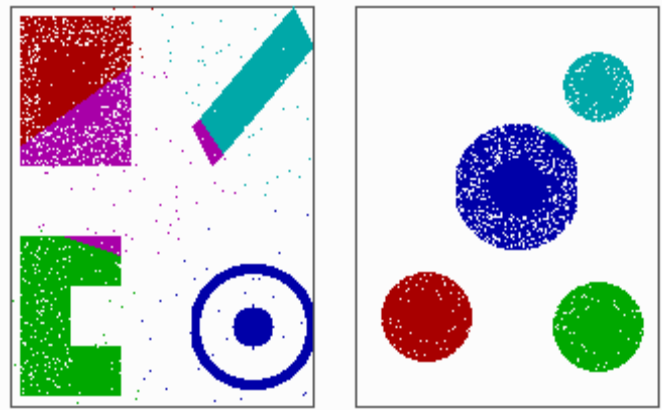

Figure 3.7: Clusters discovered by CLARANS

**Clustering results**

To assert the efficiency and effectiveness of *SCLD*, we compare it with *CLARANS*. The results are visualized by showing each cluster in a different color. To give *CLARANS* some advantage, we set the parameter *k,* which represents the number of the clusters, to 5 for DS1 and 4 for DS2. Clusters discovered by *CLARNS* are depicted in Figure 3.7. Clusters discovered by *SCLD* are depicted in figure 3.8.

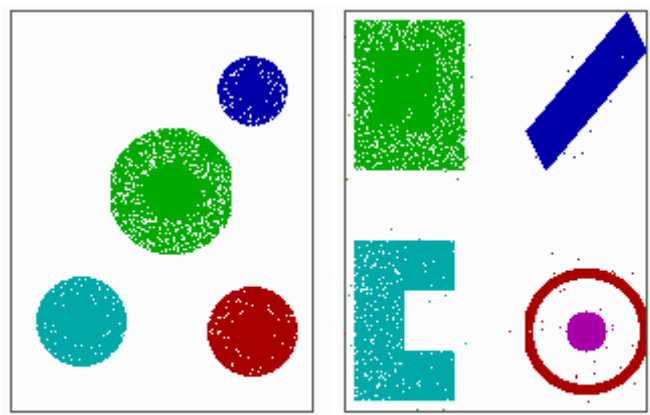

Figure 3.8: Clusterings discovered by SCLD

*SCLD* discovers all clusters (according to definition 5) and detects all noise (according to definition 6). However, *CLARANS* failed to discover clusters of arbitrary shapes. Furthermore *CLARANS* can not deal with noise.

| Number of points | *SCLD* | *CLARANS* |
|---|---|---|
| 20200 | 0.28 | 49 |
| 40200 | 0.55 | 72.28 |
| 60200 | 0.71 | 120.22 |
| 80200 | 0.99 | 352.52 |
| 100200 | 1.21 | 467.20 |
| 120200 | 1.43 | 643.40 |

**Table 3.3: run time in seconds**

The run time comparison of *SCLD* and *CLARANS* on these databases is shown in table3.3.

The results of our experiments show that the run time of *SCLD* is linear in the number of points. The runtime of *CLARANS*, however, is close to quadratic in the number of the points.

### 3.6  Conclusion

In this chapter, we introduced a grid-based clustering algorithm. We first introduced some notions that are used in the algorithm. In section 3.2, we presented the details of the algorithm. In section 3.3, we introduced an illustrative example. Section 3.4 presented the computational complexity of the algorithm. An experimental evaluation of the proposed algorithm using synthetic data was presented in Section 3.5.

We believe that The *SCLD* algorithm has the following advantages over the previous work:

1. It handles outliers or noise. Outliers refer to spatial objects, which are not contained in any clusters and should be discarded during the mining process. *CLARANS,* for example, can not handle outliers.

2. It requires only two parameters, the number of the cells in the grid structure a percentage $h$ used as in Definition 2. Instead of specifying the number of the desired clusters beforehand as input (as in *CLARANS* ), the *SCLD* algorithm finds the natural number of clusters in the spatial area.

3. When the data is updated, we do not need to re-compute all the information in the grid structure. Instead, we can do an incremental update. This is done by re-compute the information(the number of points and the mean point ) of cells that included the update. The re-compute the clusters by a breadth-first search on the cells of the grid structure.

4. It discovers clusters of arbitrary shape and it is efficient even for large spatial databases. However, *CLARANS* failed to discover clusters of arbitrary shapes. Furthermore *CLARANS* can not deal with noise.

5. Its computational complexity is much less than that of *CLARANS*.

# Chapter 4

# Clustering with Obstacles in Spatial Databases

In this chapter we discuss the problem of clustering in existence of obstacles. We formalize the notion of "clusters" in a spatial database with existence of obstacles. Then we describe how to compute the obstructed distance between two points. Then, we introduce two proposed algorithms, *CPO-WCC(Clustering in Presence of Obstacles with Computed number of cells)*,and *CPO-WFC(Clustering in Presence of Obstacles with Fixed number of cells)* to discover clusters in a spatial database with obstacles. We discuss the computational complexity of each of them. We also introduce an experimental evaluation of the effectiveness and the efficiency of the *CPO-WFC* algorithm using synthetic data. Finally, we concludes the chapter.

## 4.1 The problem

Many clustering methods have been proposed. Most of these algorithms, however, do not allow users to specify real life constraints such as the existence of physical obstacles, like mountains and rivers. Existence of such obstacles could substantially

<em>Chapter 4. Clustering with Obstacles in Spatial Databases</em>

affect the result of a clustering algorithm. For example, consider a telephone-company that wishes to locate a suitable number of telephone cabinets in the area shown in Figure 4.1 to serve the customers who are represented by points in the figure. There are natural obstacles in the area and they should not be ignored. Ignoring these obstacles will result in clusters like those in Figure 4.2, which are obviously inappropriate. Cluster $cl_1$ is split by a river, and customers on one side of the river have to travel a long way to reach the telephone cabinet at the other side. Therefore in order to generate appropriate clustering, the ability to handle such real life constraints in a clustering algorithm is important.

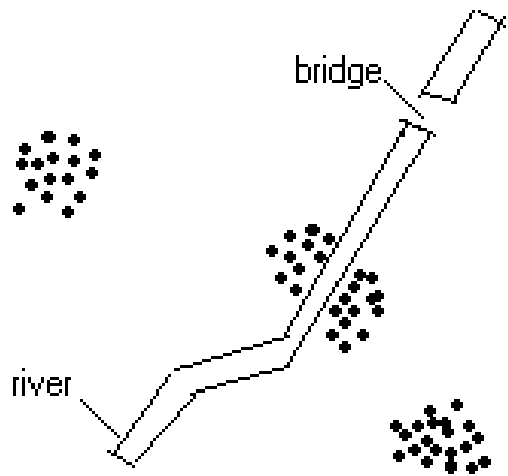

Figure 4.1 :Customers' locations and obstacles

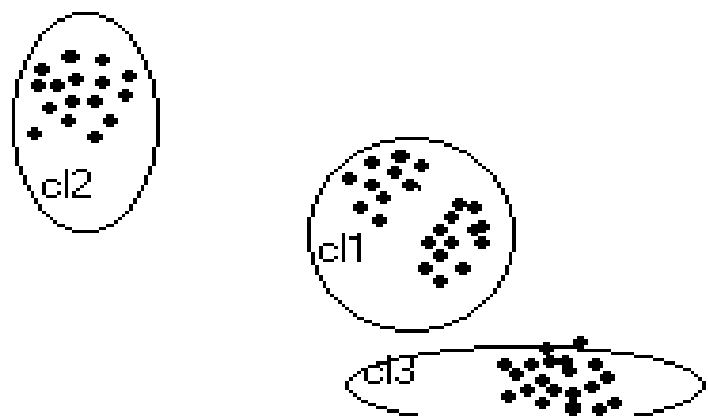

Figure 4.2: Clusters formed when ignoring obstacles

*Chapter 4. Clustering with Obstacles in Spatial Databases*

In a recent paper, [THH01], the problem of clustering with existing obstacles is investigated. [THH01] introduces an algorithm, *COD-CLARANS*, to solve the problem. Although *COD-CLARANS* generates good clustering results, there are several major problems with this algorithm. First, since the randomized search is used to determine initial centers for the clusters, the quality of the results cannot be guaranteed when the number of points, *N*, is large. Second, *COD-CLARANS* takes as an input the number of the desired clusters and the maximum number of tries the center is refined, but both numbers are generally not easy to determine in realistic applications. Third, *COD-CLARANS* can not handle outliers. Forth, when data is updated, we need to run the algorithm from scratch.

In this chapter, we propose two different efficient spatial clustering algorithms, which consider the presence of obstacles.

The proposed algorithms have several advantages over other work [THH01].
1. Handles outliers. Outliers refer to spatial objects, which are not contained in any cluster and should be discarded during the mining process.
2. Do not use any randomized search.
3. Instead of specifying the number of desired clusters beforehand, they find the natural number of clusters in the area.
4. When the data is updated, we do not need to re-compute all information in the cell grid. Instead, only information of affected cells are recomputed.

## 4.2   Basic Definitions

In the following, we formalize the necessary notions for our proposed algorithm.

**Definition 4.1:** (obstructed cell) Given a grid structure of a spatial area *S* and a set of polygon obstacles each in the form a set of vertices, a cell in the grid structure is said to be an *obstructed cell* if it is intersected by any of the obstacles otherwise the cell is said to be *non-obstructed*.
For example the top-left corner cell in figure 4.3 is obstructed.

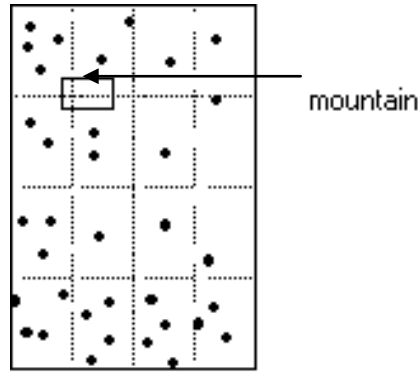

Figure 4.3 The grid structure of a spatial area

**Definition 4.2:** (dense sub-cell) Given a grid structure of a spatial area, *S*, an obstructed cell, *c* in *S*, if we construct a grid structure to the cell *c* in the same way as we do for *S*, then a **sub-cell**, *sc*, of *c* is defined as a maximal connected region of the *non-obstructed* pieces(correspond to cells in the grid structure of spatial area) in the grid structure of *c*. A sub-cell, *sc*, is said to be ***dense*** if $\frac{n_{sc}}{t} \geq p_{sc}$, where *t* is the threshold above which a cell is said to be dense and $P_{sc}$ is the percentage of the area covered by *sc* from *c*. Otherwise, *sc* is said to be non-dense.

**Definition 4.3:** (cluster) A ***cluster*** in a spatial area is a dense, maximal, connected, region of *non-obstructed* cells or sub-cells

### 4.3 The obstructed distance

The obstructed distance[THH01] between two points is the shortest Euclidean path between the two points without going through any obstacle. Our aim here is to describe how such a distance is computed[THH01]. We first define BSP-tree, which is a data structure that is used to construct a visibility graph for the vertices of the obstacles. Then, we will show how the visibility graph is used to compute the obstructed distance.

The binary-space-partition (BSP) tree [SG97] is a data structure that can efficiently determine if two points *p* and *q* are visible to each other within a region *R*. Two points *p* and *q* are visible to each other if the straight-line joining *p* and *q* dose not intersect any obstacle. To compute the obstructed distance, the BSP-tree is used to determine the set of all the visible obstacle vertices from a point *p*. This set of vertices is denoted as *vis(p)*.

**Definition 4.4** (visibility graph)[THH01] is a graph *VG=(V,G)* such that each vertex of the obstacles has a corresponding node in *V*, and two nodes $v_1$ and $v_2$ in *V* are joined by an edge in *E* iff the corresponding vertices they represent are visible to each other.

To generate the VG, [THH01] use the BSP-tree defined previously and search all other visible vertices from each vertex of the obstacles. The Visibility graph is used to find the obstructed distance between any two points in the region. The following lemma show how to use the visibility graph to compute the obstructed distance between two points. The lemma is proved in [O'R98].

**Lemma 4.1** Let *p* and *q* be two points in the region *R* and *VG=(V, G)* be the visibility graph of *R*. Let $VG' = (V', G')$ be a visibility graph created from *VG* by adding two additional nodes $p'$ and $q'$ in $V'$ representing *p* and *q*. Similar to earlier definition, $E'$ contains an edge joining two nodes in $V'$, if the points represented by two nodes are mutually visible. The shortest path between the two points *p* and *q* will be a sub-path of $VG'$.

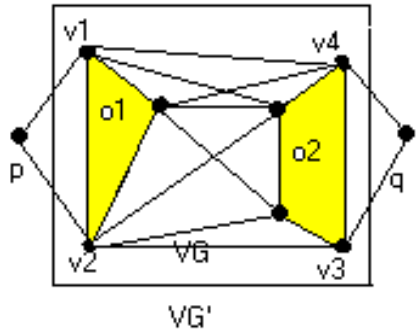

Figure 4.4  a visibility ghraph

In Figure 4.4, we show how the visibility graph *VG′* can be derived from the visibility graph *VG* of a region with two obstacles $o_1$, $o_2$. From lemma 2.1, we can see that the shortest path from *p* to *q* will begin with an edge from *p* to either $v_1$ or $v_2$, go through some path in *VG* and then end with an edge from either $v_3$ or $v_4$ to *q*. So, it is clear now that once the visibility graph, *VG*, is available, we can find the obstructed distance between two points *p* and *q* by inserting *p* and *q* into *VG* and construct *VG'* as illustrated in the above lemma. Then Dijkstra's algorithm is used for this purpose with the node *p* as the source node.

## 4.4  Algorithm *CPO-WCC*

In this section, we describe our proposed algorithm, *CPO-WCC(Clustering in Presence of Obstacles with Computed number of cells)*. The algorithm first divides the spatial area into rectangle cells of equal areas such that the average number of points in each cell is in the range from several dozens to several thousands. Then, the algorithm labels each cell as *dense* or *non-dense* according to the number of points in that cell. It labels each cell as *obstructed* (i.e. intersects any obstacle) or *non-obstructed*. Each *obstructed* cell is divided into a number of *non-obstructed* sub-cells. Again each of the new sub-cells is labeled as *dense* or *non-dense*. Then the algorithm finds the *dense* regions consisted of *non-obstructed* cells and/or sub-cells. The obtained regions are the required clusters. For each cluster the algorithm finds a center.

*Chapter 4. Clustering with Obstacles in Spatial Databases*

To find a center for a cluster the algorithm first determines the arithmetic mean of all the points in the cluster. If the mean point is not obstructed (lies in an obstructed cell), the algorithm returns it as the center of the cluster. Otherwise for each cell in the region, the algorithm determines the summation of obstructed distances [THH01] between the mean point of the cell and the mean points of the other cells in the cluster with each distance multiplied by the number of points in the other cell that contribute to this distance. The algorithm returns the mean point of the cell with the smallest summation as the center of the cluster.

Finally the algorithm outputs the clusters with their centers.

*Algorithm 4.1 CPO-WCC.*

*Input*: 1. A set of *N* objects(points) and a set of polygon obstacles in the spatial area *S*.

*Output*: clusters *S*, with a center for each cluster.

*Method*:

1. Let *La* and *Lo* be the dimensions of the spatial area. Determine two numbers, *x*, *y*, such that $\frac{x}{la} = \frac{y}{lo}$ and the average number of points in each cell, *t*, given by $\frac{N}{\frac{la}{x} * \frac{lo}{y}}$ ranges from several dozens to several thousands.

2. Divide the spatial area into $\left(\frac{la}{x} * \frac{lo}{y}\right)$ rectangular cells that have equal areas by dividing the longitude and latitude of the spatial area into ($\frac{la}{x}$) equal segments.

3. For each cell, c, we determine the following parameters:
    - $n_c$ : the number of the objects in the cell.
    - $m_c$ : the mean of the points in the cell.

4. For each cell, *c*, if $n_c \geq t$, then label *c* as *dense*, otherwise label *c* an *non-dense*.

5. For each obstacle, *O*, label all cells that intersect the boundary of *O* as *obstructed*.

6. For each *obstructed* cell, *c*, apply algorithm 4.2 to find *non-obstructed* sub-cells in *c*.

7. For each sub-cell, *sc*, obtained in step 6, if $\frac{n_{sc}}{t} \geq p_{sc}$, then label *sc* as *dense*. Otherwise label *sc* as *non-dense*.

8. For each *dense*, *non-obstructed* cell that is not, previously, processed in the current step, or *dense* sub-cell that also is not processed before in the current step, we examine its neighboring non-obstructed cells or sub-cells to see if the average density within this small area is greater than or equal *t*. If so, this area is marked and all *dense* cells or sub-cells that are just examined are put into a queue. Remove from the queue each *dense* cell or sub-cell that has been examined before in a previous iteration. Each time we take one cell or sub-cell from the queue and repeat the same procedure. When the queue is empty, we have identified one cluster, *Cl*.

9. For each cluster, *Cl*, constructed in step 8, apply algorithm 4.3 to find center for *Cl*.

10. utput the constructed clusters with their centers.

*Algorithm 4.2 Find_non-obstruced_sub-cell(c)*

**Input:** an *obstructed* cell, *c*.

**Output:** a number of *non-obstructed* sub-cells in *c*

**Method:**

1. Divide the cell *c* into a number of small pieces of equal areas in the same way as we divide the spatial area such that the average number of the points in each piece(smaller than the average number of the points in a cell inside the spatial area) is in the range from several dozens to several hundreds.

2. For each small piece, *p*, label *p* as either *obstructed* (i.e. intersects any obstacle) or *non-obstructed*.

3. For each *non-obstructed* piece, *p*, that is not marked before in this step, the area constituted from *p* and its *non-obstructed* neighbors is marked and all *non-obstructed* neighbors we just examined are put into a queue. Each time, we take one piece from the queue and repeat the same procedure except that

   those *non-obstructed* pieces that are not marked before are enqueued. When the queue is empty, we have identified one sub-cell.

4. For each sub-cell, *sc*, obtained in step 3, we determine the following parameters:

    $n_{sc}$: the number of the objects in the sub-cell.

    $m_{sc}$: the mean of the points in the sub-cell.

    $P_{sc}$: the percentage of the area covered by *sc* from *c*.

5. Output the sub-cells formed in step 3 with their parameters.

*Algorithm 4.3 Find_center(r)*

*Input:* A cluster, *cl*.

*Output*: a center for the cluster *cl*.

*Method:*

1. Calculate the mean point, *mp*, of all the points in the cluster *cl*.
2. If this mean is not in any obstacle, then return *mp*.
3. Otherwise, for each cell, *c*, with mean $m_c$, in the cluster, find *cost(c)* which is $\sum_{i \in cl} n_i (d'(m_c, m_i))^2$, where *i* is a cell in the cluster and $d'(m_c, m_i)$ is the obstructed distance, between $m_c$ and $m_i$.
4. Return the mean point of the cell with the minimum cost.

### 4.4.1 Example

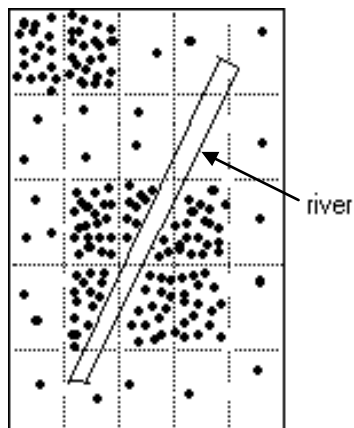

Figure 4.5: the spatial area with objects and obstacles

*Chapter 4. Clustering with Obstacles in Spatial Databases*

In this section, we present an example to apply the *CPO-WCC* algorithm. Suppose that we have a set of 5,000 objects and a set of obstacles each in the form of a polygon in a spatial area. Figure 4.5 shows a sketch of the spatial area in which objects are represented by points and the obstacle is a river represented by a polygon. Clusters are founded as follows.

1. The algorithm first finds two rationales, *x, y*, such that $\frac{x}{la} = \frac{y}{lo}$ and $\frac{5,000}{(la/x)*(lo/y)}$ is in the range from several dozens to several thousands, where *la, lo* are the latitude and longitude (the dimensions) of the spatial area. In our example *la* and *lo* equal 5 and 7.5 length units respectively.

| Cell number | Number Of points |
|---|---|
| 1 | 400 |
| 2 | 400 |
| 12 | 500 |
| 13 | 300 |
| 14 | 600 |
| 17 | 500 |
| 18 | 100 |
| 19 | 600 |

Table 4.1: The number of points in some cells

So if we take *x*=1 and *y*=1.5, *x* and *y* will satisfy the required conditions.

2. The algorithm divides the spatial area into $25(=(la/x)*(lo/y))$ equal cells by dividing the latitude and the longitude into $5(=la/x=lo/y)$ equal segments. So we obtained the dotted grid structure in Figure 4.5.

3. The cells in Figure 4.5 are numbered from left to right and from top to down. For each cell, c, the algorithm determines two parameters, $n_c$, the number of points in the cell and $m_c$, the mean of points in the cell. Table 4.1 shows the first parameter for some cells. We assume that the remaining points are distributed among the remaining cells so that no cell contain more than 100 points and that the mean of the points in every cell is in the center of the cell.

4. The algorithm then defines $t$ as $t = \dfrac{5{,}000}{25} = 200 (= \dfrac{N}{(la/x)*(lo/y)})$. For each cell, c, if $n_c \geq t$, then c is labeled as *dense*. So, cells 1, 2, 12, 14, 17 and 19 are *dense*.

5. For each obstacle, the algorithm labels all cells that intersect *O* as *obstructed*. So, among the *obstructed* cells in our example are cells 13, 14, 17 and 18.

6. For each *obstructed* cell, *c*, the algorithm finds the *non-obstructed* sub-cells of *c* as follows. To find the *non-obstructed* sub-cells of cell 13, we divide it into *p* equal pieces such that the number of the points in each piece($\dfrac{n_{13}}{p}$) is in the range from several dozens to several hundreds. If we take *p* = 24, then $\dfrac{300}{24} \approx 13$ which is in the required range. The algorithm then labels each piece as *obstructed* or *non-obstructed*. For each *non-obstructed* piece, *p*, the area constituted from *p* and its *non-obstructed* neighbors is marked and all *non-obstructed* neighbors we just examined are put into a queue. Each time, we consider one piece from the queue. All the neighbors of the piece except those that have been already examined. When the queue is empty, we have identified a portion of the obstructed cell, *c*, as a non-obstructed portion of *c* .

   For example, Figure 4.6 shows an *obstructed* cell, cell $c_{13}$ in Figure 4.5, where the red and blue area represent the non-obstructed portions of cell $c_{13}$.

For each sub-cell, sc, the algorithm determines parameters $n_{sc}$, the number of points in the sub-cell , $p_{sc}$, the percentage of the sub-cell area to the total area of the cell and $m_{sc}$, the mean of points in this sub-cell. For example $n_{sc1}=125$, $Psc_1=0.25$, $n_{sc2} = 150$ and $P_{sc2}= 0.25$. For each sub-cell, sc, if $\frac{n_{sc}}{t} \geq P_{sc}$, then sc is labeled as *dense*. Otherwise sc is labeled as *non-dense*. So both $sc_1$ and $sc_2$ are *dense* sub-cells.

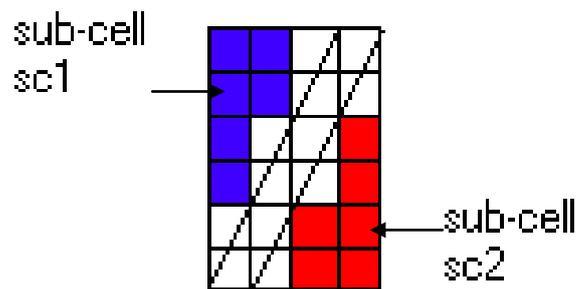

Figure 4.6 : the non-obstructed sub-cells of an obstructed cell

7. We obtain the clusters as shown in Figure 4.7.

8. The algorithm then determines center for clusters as follows. For each cluster calculate the mean, *m*, of the points in the *dense* cells or sub-cells that constitute the cluster. If this mean is not in any obstacle, then *m* is the cluster center. Otherwise, for each cell, *c*, in the cluster, find

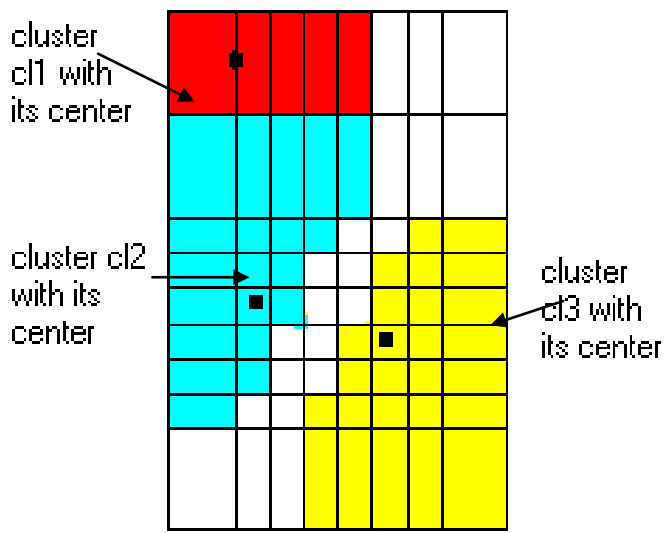

Figure 4.7: the final results of CPO-WCC

cost(*c*) which equals $\sum_{d \in R}(d'(m_c, m_d))^2$, where $d'(m_c, m_d)$ is the *obstructed* distance which defined as the shortest Euclidean path from $m_c$ to $m_d$ *without* cutting through any obstacles. Computing the obstructed distance was discussed in Section 4.3. The summation is taken over all cells that constitute the cluster. The required center is the mean of the cell with the minimum cost. By doing so, we obtain the cluster centers as shown in Figure 4.7

### 4.4.2  Analysis of the *CPO-WCC* algorithm

We analysis the complexity of the *CPO-WCC* in this section. The following notations are defined for this discussion:

*N*: the number of data objects in the database *D*.

*m*: the number of cells in the spatial area, if we divide the spatial area such that the number of points in each cell is in the range from several dozens to several thousands , *m<<N*.

*m'*: the number of cells in the spatial area, if we divide the spatial area such that the number of points in each cell is in the range from several dozens to several hundreds, *m<m'<<N*.

*G*: the visibility graph.

*|V|*: the number of vertices in *G*.

*|E|*: the number of the edges in *E*.

Steps 1 and 2 take a constant time. Step 3 scans the database once so it takes *O(N)* time. In the worst case, steps 4, and 5, may need to scan all cells, so they take *O(m)*. Accordingly, the running time for the first five steps is *O(m)+O(m)+O(N)*. Since *m<<N* the complexity is *O(N)* time.

In the worst case, i.e. all cells may be obstructed, step 6 will be applied on all cells. But this is equivalent to re-execute steps 1 and 2 on the spatial area such that the number of points in every cell in the new grid is in the range from several dozens to several hundreds. The number of the cells in the new grid is m'. Afterwards, to go through step 6, we need to scan all cells in the new grid to label them as *obstructed* or *non-obstructed*, form the *non-obstructed* sub-cells and compute the parameter *p* for each sub-cell and this scan takes *O(m')* time and then scan the data points again to determine the other parameters to each sub-cell which takes *O(N)* time. So, totally step 6 takes *O(N)+O(m')* time. In step 7, we need to scan all the formed sub-cells to label them as *dense* or *non-dense* but this takes *O(m')* time. Step 8 in the worst case takes *O(m')* time.

The running time of step 9 is $O(m|V|^2)+O(m^2/|V|)$. The worst case is when the mean point of the points in the cluster is inside an obstacle, we will need to scan all cells in the cluster and find the cell whose mean point is the nearest to the mean points of all the other cells in the cluster. For each cell, to determine the sum of the obstructed distances between it's mean point and the mean points of the other cells in the cluster, this takes $O(|V|^2)+O(m/|V|)$. Calculation of this sum includes the following to operations. First, for the mean point of this cell the Dijkstra algorithm[O'R98] is applied which takes $O(|V|^2)$ time. Second, for each of the cells in the cluster, we need to look up its visibility information. This takes *O(|V|)* to each cell, so the second operation takes as a total *O(m/|V|)*. Now, since the calculation of summation is needed for each cell in the cluster, the total complexity of this step is $O(m|V|^2)+O(m^2/|V|)$.

The total complexity of the *CPO-WCC* algorithm is the sum of the running time of all the steps:

$$O(CPO\text{-}WCC) = O(N) + O(N) + O(m') + O(m') + O(m') + O(m|V|^2) + O(m^2/|V|)$$
$$= O(N) + O(m|V|^2) + O(m^2/|V|).$$

## 4.5 Algorithm *CPO-WFC*

In this section, we describe the *CPO-WFC(Clustering in Presence of Obstacles with Fixed number of cells)* algorithm. The *CPO-WFC* differs from the *CPO-WCC* in two main matters. First, *CPO-WCC* determines the number of the cells in the grid by itself while *CPO-WFC* takes this number as an input. Second, the two different techniques find the clusters using two completely different searching methods. These methods are illustrated in detail in its technique. The algorithm first divides the spatial area into *m,* which is an input*,* rectangular cells of equal areas. Then, the algorithm labels each cell as *dense* or *non-dense* (according to the number of points in that cell and an input threshold) also *obstructed* (intersects an obstacle) or *non-obstructed*. For each *obstructed* cell, the algorithm finds *non-obstructed* sub-cells by dividing the cell into *m* pieces of equal areas and finding the maximal, connected areas of *non-obstructed* pieces as the required *non-obstructed* sub-cells of that cell.

The algorithm also determines which of those sub-cells are *dense* by the same way as for cells. The algorithm finds the maximal, connected, regions of *dense, non-obstructed* cells or *dense* sub-cells as the required clusters. The algorithm finds a center for each obtained cluster.

for each *non-empty*, *non-dense non-obstructed* cell or *non-empty*, *non-dense* sub-cell, the algorithm assigns the cell or the sub-cell to the neighboring cluster that satisfy the following conditions. First, the addition of the cell or the sub-cell to the cluster dose not violate the first condition in the definition of the cluster (see Definition 3.6). Second, if there are more than one cluster that satisfy the first condition, the cell or the sub-cell is added to the one whose center in the nearest (using the obstructed distance [THH01]) to the mean point of the cell or the sub-cell. Finally the algorithm re-computes a center for each extended cluster and outputs the obtained centers as the centers of the clusters in the spatial area.

To find a center for a cluster the algorithm first determines the arithmetic mean of all the points in the cluster. If the mean point is not obstructed (lies in an obstructed cell), the algorithm returns it as the center of the cluster. Otherwise for each cell in the region, the algorithm determines the summation of obstructed distances [THH01] between the mean point of the cell and the mean points of the other cells in the cluster with each distance multiplied by the number of points in the other cell that contribute to this distance. The algorithm returns the mean point of the cell with the smallest summation as the center of the cluster.

*Algorithm 4.4 CPO-WFC.*

*Input*:

1. A set of *N* objects and a set of polygon obstacles in a spatial area *S*.

2. A square number, *m*, which represents the number of cells in the spatial area such that $m<<N$.
3. A percentage, *h*, to be used in determining the dense cells according to definition 3.2

**Output**: clusters in the spatial area *S* with their centers.

**Method**:

1. Let $w = \sqrt{m}$.
2. Divide the spatial area into *m* rectangular cells that have equal areas by dividing each of the dimensions of the spatial area into *w* equal segments.
3. For each cell, c, we determine the following parameters:
    - $n_c$ : the number of the points in the cell.
    - $m_c$ : the mean point of all points in the cell.
4. Let $d = round(\frac{N}{m} * h)$.
5. For each cell, c, if $n_c \geq d$, then c is labeled as *dense*, otherwise c is labeled as *non-dense*.
6. For each obstacle, *O*, label all cells that intersect the boundary of *O* as *obstructed*.
7. For each *obstructed*, non-empty cell, *c*, apply algorithm 4.5 to find *non-obstructed* sub-cells in *c*.
8. For each *dense, non-obstructed* cell or *dense* sub-cell, *c*, which is not processed before, construct a new cluster, *cl*, and mark the cell or sub-cell *c* as an element of *cl*. Put the *dense*, *non-obstructed*, neighboring cells and the *dense* neighboring sub-cells of *c* in a list, *Q*. While *Q* isn't empty, take the first element, *c'*, from *Q*, mark *c'* as an element of the cluster *cl* and add the *dense*, *non-obstructed*, neighboring cells and the *dense* neighboring sub-cells of *c'* to *Q*.
9. For each cluster, *cl*, constructed in step 8, apply algorithm 4.3 to find a center for *cl*.
10. Put all non-empty, *non-dense*, *non-obstructed* cells and non-empty, *non-dense* sub-cells, that satisfy the condition that each such cell or sub-cell is a neighbor to a cluster, in a queue, *QN*, sorted in a descending order according to the number of the data points in each of them.

11. While *QN* isn't empty
    11.1 Take the first element, *e*, of *QN*.
    11.2 Find all neighboring clusters to *e*.
    11.3 Determine which of those clusters will not violate the first condition in the definition of cluster(Definition 3.2) after the addition of *e*.
    11.4 Assign e to the one of the clusters founded in step 11.3 that has the nearest (using obstructed distance [THH01]) center to the mean point of e.
12. Re-evaluate a center for each extended cluster by applying algorithm 4.3
13. Output the obtained clusters with their centers.

*Algorithm 4.5 Find_non-obstruced_sub-cell(c)*

**Input:** an *obstructed* cell, *c*.

**Output:** a number of *non-obstructed* sub-cells in *c*

**Method:**

1. Divide the cell, *c*, into *m* rectangular pieces that have equal areas by dividing each of the dimensions of the cell into *w* equal segments.
2. For each obstacle, *O*, that interests *c*, all pieces in *c* that intersects the boundary of *O* are labeled as obstructed.
3. For each *non-obstructed* piece, *p*, which is not processed before, construct a new sub-cell, *sc*, and mark the piece, *p*, as an element of *sc*. Put the *non-obstructed*, neighboring pieces of *p* in a queue, *Q*. While *Q* isn't empty, take the first element, *p'*, from *Q*, mark *p'* as an element of the sub-cell *sc* and add the *non-obstructed*, neighboring pieces of *p'* to *Q*.
4. For each sub-cell, *sc*, obtained in step 3, we determine the following parameters:
   $n_{sc}$: the number of the objects in the sub-cell.
   $m_{sc}$: the mean of the points in the sub-cell.
   $P_{sc}$: the percentage of the area covered by *sc* from *c*.
5. For each obtained sub-cell, sc, if $\frac{n_{sc}}{d} \geq p_{sc}$, then label *sc* as *dense*, otherwise label *sc* as *non-dense*.
6. Output the obtained sub-cells.

**4.5.1 Example**

In this section, we describe our algorithm by an example. Suppose that we have a set of 5,000 objects and a set of one polygon obstacle in a given spatial area. Figure 4.8 shows a sketch of the spatial area in which the points represent the objects and the river represents the obstacle. And we are asked to find clusters with their centers in the spatial area given that $m = 36$ and $h = 0.9$. The algorithm works as follow.

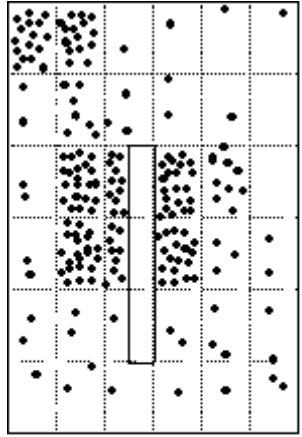

Figure 4.8: the spatial area with objects and obstacles

1. $w = 6$ ( $w = \sqrt{m}$ ).
2. The spatial area is divided into 36 equal cells by dividing each of the spatial area dimensions into 6 ($= \sqrt{36}$) equal segments. So we obtained the dotted grid structure in Figure 4.8. The cells in Figure 4.8 are numbered from left to write and from top to down.
3. For each cell, *c*, the algorithm determines the two parameters $n_c$, $m_c$ described in the algorithm. Table 4.2 shows the first parameter for some cells. We assume that the remaining points are distributed among the remaining cells so that no cell contain more than 100 points and that the mean of the points in every non-empty, *non-obstructed* cell is in the center of the cell.
4. $d = 125$ ( $d = round(\frac{N}{m} * h)$ ).
5. Each cell, c, with $n_c \geq d$ is labeled as *dense*. So, cells 1, 2, 14, 16, 20 and 22 are the only *dense* cells.

6. The cells that intersect the river are marked as *obstructed* cells. Using simple geometrical concepts does this. So cells 15, 21 and 27 are labeled as *obstructed* cells.

| Cell number | Number Of points |
|---|---|
| 1 | 400 |
| 2 | 400 |
| 14 | 500 |
| 15 | 300 |
| 16 | 600 |
| 20 | 180 |
| 21 | 100 |
| 22 | 600 |

Table 4.2: The numbe of points in some cells

7. Algorithm 3.2 is applied on cells 15, 21 and 27 to find *non-obstructed* sub-cells from each of them. For example to find *non-obstructed* sub-cells from cell 15 the following is done.

    7.1. The cell is divided into 36 equal pieces by the same way as we divided the spatial area.

    7.2. The pieces that intersects the river are marked as *obstructed* pieces. Using simple geometrical concepts does this.

    7.3. The maximal connected regions (sub-cells) of the *non-obstructed* pieces are obtained from the pieces of cell by a breadth-first search. Figure 4.9 illustrates the obtained sub-cell from cell 15. By similar way the algorithm finds the *non-obstructed* sub-cells from cells 9 and 21.

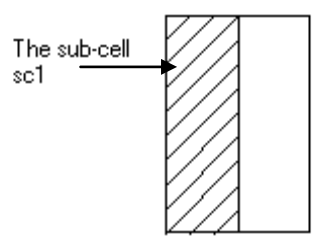

Figure 4.9: cell 15 with the sub-cell, sc1, obtained by applying algorithm 3.5 on it 7.4. The parameters $n_{sc}$, $m_{sc}$, $p_{sc}$ defined above in the algorithm are determined for each obtained sub-cell. For example, for sub-cell $sc_1$ of cell 15 $n_{sc_1}$ =300 and $p_{sc_1}$ =0.5.

7.5. The algorithm determines which of these sub-cells are dense as follows. For example the sub-cell $sc_1$ is dense because $\frac{n_{sc_1}}{d}$ which equals $\frac{300}{125}$ is greater than $p_{sc_1}$ which equals 0.5.

8. The maximal, connected regions (clusters) of the *non-obstructed*, *dense* cells or the *dense* sub-cells are determined by a breadth-first search. Figure 4.10 illustrates such clusters in the given spatial area with each cluster in a different color.

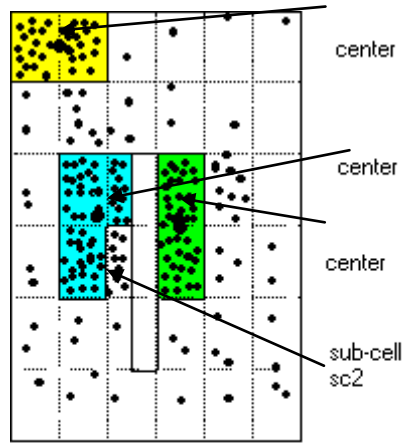

Figure 4.10: the resulted regions after step 8 of the algorithm with the centers obtined in step 9.

9. For each cluster constructed in step 8, algorithm 3.3 is applied to find a center to that cluster. Algorithm 3.2 returns the arithmetic mean of the points in each cluster as its center. Those centers are shown in Figure 4.10.

10. Suppose that sub-cell, $sc_2$, of cell 8 is the first element in *QN* of our example.

11. We begin trying to add $sc_2$ to the neighboring cluster to $sc_2$ with the nearest center. But it is clear that there is only one of such cluster and the addition of $sc_2$ to it will not violate the first condition in the definition of the cluster(Definition 3.2). So $sc_2$ is added to that cluster. The same is repeated to each element of *QN*. Figure 4.11 shows the resulting clusters after step 11 with a new center for each extended cluster.

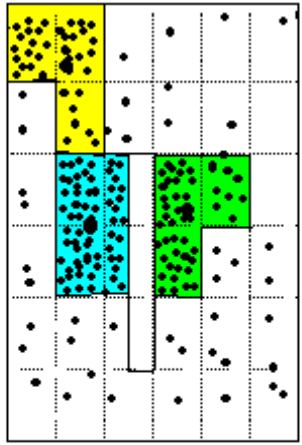

Figure 4.11: the resulted centers of the algorithm

### 4.5.2   Analysis of the *CPO-WFC* Algorithm

We analysis the complexity of the *CPO-WFC* in this section. The following notations are defined for this discussion:

*N*: the number of data objects in database *D*.

*m*: the number of the cells in the spatial area, $m<<N$.

*G*: the visibility graph.

$|V|$: the number of vertices in *G*.

$|E|$: the number of the edges in *E*.

Steps 1, 2, and 4 take a constant time because they include simple calculations. Step 3 scans the database only once so it takes $O(N)$ time. In the worst case, steps 5, and 6 , may need to scan all cells, so they take $O(m)$. Accordingly, the running time for steps 5 and 6 $O(m)+O(m)= O(m)$ time. Step 7 takes time of $O(N)+O(m^2)$. The expected running time of step 8 is $O(m)$. Since $m<<N,$ The running time of the first 8 steps is $O(N)+O(m^2)$.

The running time of step 9 is $O(m|V|^2)+O(m^2|V|)+O(|V|)$. This is because in the worst case, which occurs when the mean point of the points in a region is inside an obstacle, we will need to scan all cells in the region and find the cell whose mean point is the nearest to the mean points of all the other cells in the region. For each cell, to determine the sum of the obstructed distances between it's mean point and the mean points of the other cells in the region, this takes $O(|V|^2)+O(m|V|)$[J99]. Since the calculation of summation is needed for each cell in the region, the total complexity of doing this is $O(m|V|^2)+O(m^2|V|)$ and the Ray Crossings algorithm[O'R98] takes $O(|V|)$ to determine whether the mean point is inside an obstacle.

The expected running time of step 10 is $mlog(m)$. Step 11 takes time of $O(m)$. The complexity of step 12 is like step 9.

The total complexity of CPO-WFC algorithm is the sum of the running time of all the steps:

$$O(CPO) = O(N) + O(m^2) + O(m|V|^2) + O(m^2|V|) + O(|V|) + mlog(m) + O(m)$$

$$= O(N) + O(m|V|^2) + O(m^2|V|).$$

### 4.5.3 Marking the obstructed cells and determining whether a point is inside an obstacle

In this section, we describe how the algorithm marks the cells that intersect the boundary of an obstacle as obstructed. We also describe how can the algorithm determines whether a point is inside an obstacle. Given an obstacle and the minimum length, say $e$, of the two dimensions of a cell in the grid structure, the algorithm marks the cells that intersect the boundary of an obstacle as follows. For each boundary edge of the obstacle, the algorithm marks the cells that intersect the vertices of the obstacle that determine that edge as obstructed. Then the algorithm divides this edge by the mean point of the two vertices that determine it and marks the cell that contains the mean point as obstructed and repeat the same procedure with the two segments that constituted by the mean point until the length of each of them is less than or equal $e$.

While we calculate a center for a cluster we need to check whether a point is inside an obstacle. For doing that our algorithm applies the Ray Crossings algorithm [O'R98], which takes $O(t)$ to determine whether a point is inside an obstacle of $t$ vertices.

## 4.6 Performance Study

In this section, we will have a look at the performance of the *CPO-WFC* algorithm by performing experiments on a PC with a Pentium 900 Mhz processor and a Western digital 750 rpm hard disk. We use two synthetic databases, *DS1* and *DS2*, for these experiments. Those databases are shown in Figures 4.12 and 4.13. In *DS1*, there are four clusters of different size. Three of them are balled-shaped and the fourth one is a rectangular one which is surrounded by a polygon obstacle. In *DS2*, there are four ball-shaped clusters of significantly differing sizes. One of them is split by a polygon obstacle into two clusters. *DS2* contains additional noise. The number of data points in DS1 is 40000 data point and 50200 in *DS2*.

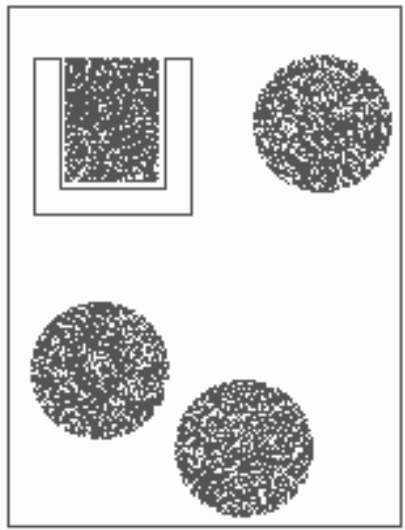

Figure 4.12: DS1

The experiments processed as follows. First, we assess the efficiency and effectiveness of the *CPO-WFC* by running it on *DS2* for different values of the parameter *m* which represents the number of the formed cells in the spatial area.

Second, we compare *CPO2* with the performance of *COD-CLARANS* [HTT01] which is the only clustering work that consider the presence of the obstacles in the spatial area.

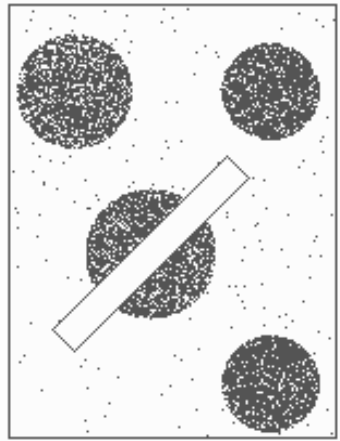

Figure 4.13: DS2

**Varying *m***

In this experiment, we vary the number of cells that are generated for *DS2* by tuning the parameter *m*. There are two reasons in doing this. First, because more accuracy will be granted when *m* is increased, we like to investigate how the quality of the clusters is affected by the choice of *m*. Second, we also like to explore how the choice of *m* will affect the running time of *CPO-WFC*.

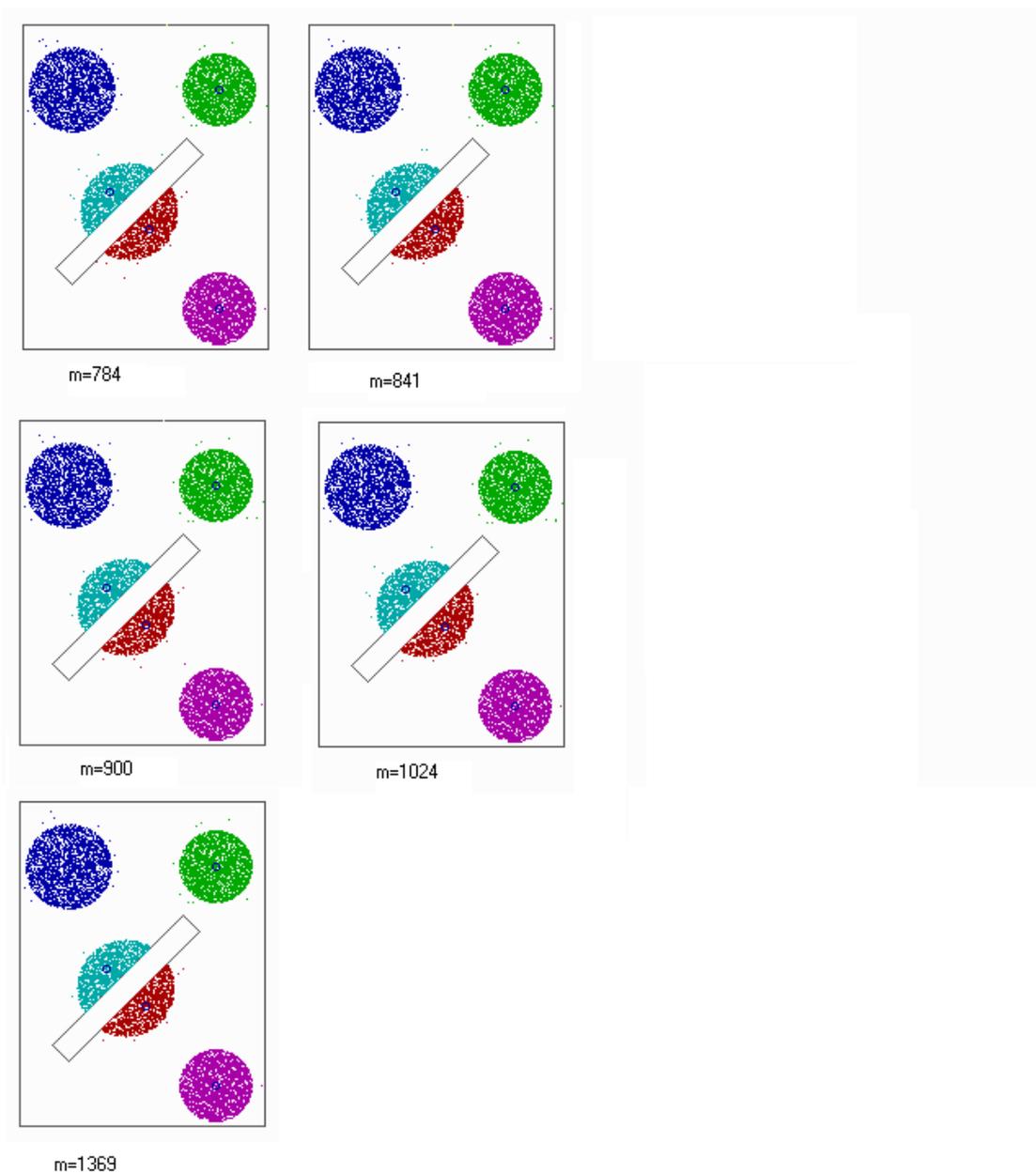

Figure 14.4 : The clustering results for CPO-WFC on DS 2 for different values of m

To show the results of the clustering process, we visualize each cluster by a different color. For the first reason, Figure 4.14 shows the clustering results for running *CPO-WFC* on *DS2 for* different values for *m*. From the figure, we note that as we increase *m* the quality of the clustering process is increased by eliminating more noise points. For the second purpose, table 4.3 shows the running in seconds for running *CPO-WFC* on *DS2* for different values of *m*.

| *m* | Running time in seconds |
|---|---|
| 676 | 4.82 |
| 841 | 5.25 |
| 1024 | 5.96 |
| 1156 | 6.41 |
| 1296 | 7.1 |
| 1369 | 7.9 |

**Table4.3:The running time for different values of *m***

From the table, we note the time does not increase dramatically as *m* increases.

**Clustering results**

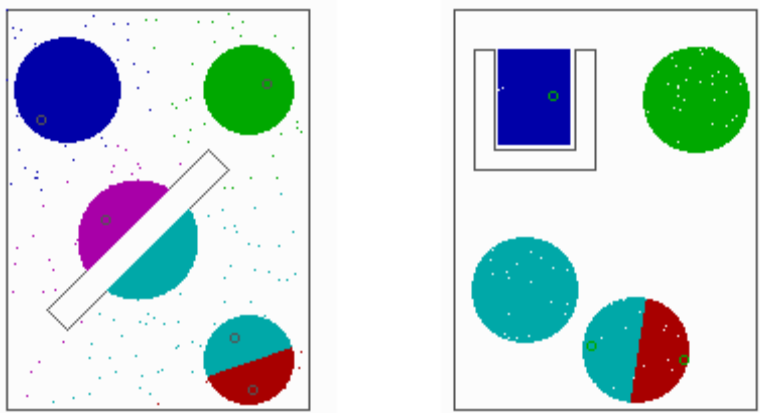

Figure 4.15: The clusters discovered by COD-CLARANS

To assert the efficiency and effectiveness of CPO2, we compare it with *COD-CLARANS*. The clustering results are visualized by showing each of them in a different color. To give *COD-CLARANS* some advantage, we set the parameter *k*, to 4 for *DS1* and 5 for *DS2*. The clusters discovered by *COD-CLARNS* are depicted in figure 4.15. The clusters discovered by *CPO-WFC* are depicted in figure 4.16.

*CPO-WFC* discovers all clusters and detects all noise. However, *COD-CLARANS* failed to discover clusters of arbitrary shapes. Furthermore *COD-CLARANS* can not deal with noise.

| Number of points | CPO-WFC | COD-CLARANS |
|---|---|---|
| 20200 | 2.31 | 73.5 |
| 40200 | 4.53 | 108.42 |
| 60200 | 5.85 | 180.33 |
| 80200 | 8.16 | 528.78 |
| 100200 | 9.97 | 700.8 |
| 120200 | 11.79 | 965.1 |

**Table 3.4: run time in seconds**

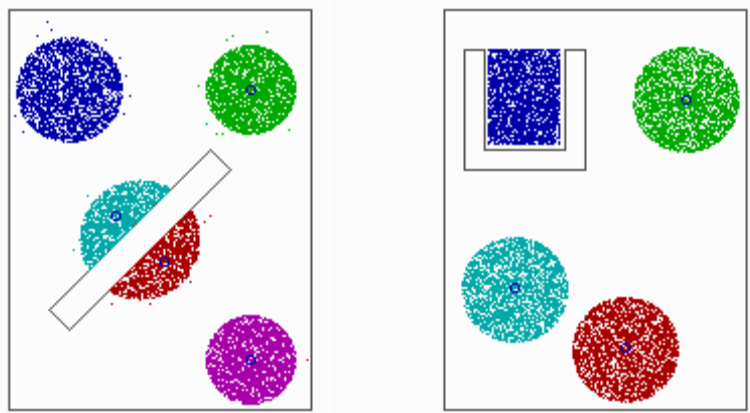

Figure 4.16: The clusters discovered by CPO-WFC

The run time comparison of *CPO-WFC* and *COD-CLARANS* on these databases is shown in table 4.4.

The results of our experiments show that the run time of *CPO-WFC* is linear in the number of points. The runtime of *COD-CLARANS*, however, is close to quadratic in the number of the points.

## 4.7 Conclusion

In this chapter, we introduced two grid-based clustering algorithm that consider the existence of physical obstacles. We first illustrated the problem that arise due to ignoring physical obstacles. Section 4.2 explained how the obstructed distance is computed. We also introduced some notions that are used in the algorithms in section 4.3. In sections 4.4, 4.5 we presented the details of the proposed algorithms, CPO-WCC and CPO-WFC respectively. An experimental evaluation of the proposed algorithm, CPO-WFC, using synthetic data was presented in Section 4.6.

We believe that The CPO-WFC and CPO-WCC algorithms have the following advantages over the previous work[THH01]:

1. Handles outliers.
2. Do not use any randomized search.
3. Instead of specifying the number of desired clusters beforehand, they find the natural number of clusters in the area.
4. When the data is updated, we do not need to re-compute all information in the cell grid. Instead, only information of affected cells are recomputed.

# Chapter 5

# Conclusion and Future Work

In this thesis, we proposed an efficient algorithm for spatial clustering of large spatial databases. The algorithm overcomes the problems of the previous work. The algorithm divides the spatial area into rectangular cells and labels each cell as *dense* (contains relatively large number of points) or *non-dense*. The algorithm finds all the maximal, connected, dense regions that form the clusters by a breadth-first search and determine a center for each region.

The algorithm has the following advantages over the previous work
1. It handles outliers or noise.
2. It requires only two input parameters, the data records that we want to clustered, the number of the cells. Unlike some other clustering algorithms [23], the proposed algorithm does not require determining the number of the desired clusters in advance. In stead, the natural clusters existing in the spatial database are obtained.      .

3. When the data is updated, we do not need to re-compute all the information in the grid structure. Instead, we can do an incremental update.

4.  It discovers clusters of arbitrary shapes.

5. Its computational complexity is comparable to the complexity spatial clustering algorithms, *$O(n^2)$* [23], *$O(nlogn)$* [11].

We also implemented the proposed algorithm and compare it with CLARANS algorithm [NH94]. The experiments showed that our algorithm is more superior in both running time and accuracy of the results.

The second contribution of this work is introducing efficient algorithms of spatial clustering with presence of obstacles. In [29], the problem of clustering with presence of obstacles is introduced.

 The  proposed  algorithms  are  called  as *CPO-WCC (Clustering in Presence of Obstacles with Computed number of cells)*, and *CPO-WFC (Clustering in Presence of Obstacles with Fixed number of cells).*

The *CPO-WCC* algorithm first divides the spatial area into rectangle cells of equal areas such that the average number of points in each cell is in the range from several dozens to several thousands. Then, the algorithm labels each cell as *dense* or *non-dense* (according to the number of points in that cell) also *obstructed* (i.e. intersects any obstacle) or *non-obstructed*. Each *obstructed* cell is divided into a number of *non-obstructed* sub-cells. Again each of the new sub-cells is labeled as *dense* or *non-dense* (according to the number of point in this sub-cell. Then the algorithm finds the *dense* regions of *non-obstructed* cells or sub-cells. The obtained regions are the required clusters. For each cluster the algorithm finds a center. Finally the algorithm outputs the clusters with their centers.

The *CPO-WFC* algorithm first divides the spatial area into *m,* which is an input, rectangular cells of equal areas. Then, the algorithm labels each cell as *dense* or *non-dense* (according to the number of points in that cell and an input threshold) also *obstructed* (intersects an obstacle) or *non-obstructed*. For each *obstructed* cell, the algorithm finds *non-obstructed* sub-cells by dividing the cell into *m* pieces of equal areas and finding the maximal, connected areas of *non-obstructed* pieces as the required *non-obstructed* sub-cells of that cell.

The algorithm also determines which of those sub-cells are *dense* by the same way as for cells. The algorithm finds the maximal, connected, regions of *dense, non-obstructed* cells or *dense* sub-cells. The algorithm finds a center for each obtained region.

For each *non-dense*, *non-obstructed* cell or *non-dense* sub-cell the algorithm assign it to the region that satisfy the following three conditions, if such a region exists. First, the cell or sub-cell is a neighbor to one of the cells or the sub-cells in that region. Second, the addition of that cell or sub-cell to the region won't degrade the overall density of the newly resulting region to be below the least possible density of a *dense* cell. Third the center of that region is the nearest (using the obstructed distance [29]) to the mean point of the cell or sub-cell among all regions satisfying the first and the second conditions. Finally the algorithm re-evaluates a center for each extended region and outputs the obtained centers as the centers of the relatively dense regions in the spatial area.

To find a center for a region the algorithm first determines the arithmetic mean of all the points in the region. If that mean point is not obstructed (lies in an obstructed cell), the algorithm returns it as the center of the region. Otherwise for each cell in the region, the algorithm determines the summation of obstructed distances [29] between the mean point of the cell and the mean points of the other cells in the region with each distance multiplied by the number of points in the other cell that contribute to this distance. The algorithm returns the mean point of the cell with the smallest summation as the center of the region.

We also implemented *CPO-WFC* and compare it with COD-CLARANS algorithm [29]. The experiments showed that *CPO-WFC* is more superior in both running time and accuracy of the results.

# CHAPTER 5. CONCLUSSION AND FUTURE WORK

The proposed algorithms, *CPO-WCC*, and *CPO-WFC* have several advantages over other work in [29].

1. They handle outliers or noise. Outliers refer to spatial objects, which are not contained in any cluster and should be discarded during the mining process.

2. They dose not use any randomized search.

3. Instead of specifying the number of desired clusters beforehand, They find the natural number of clusters in the area.

4. When the data is updated, we do not need to re-compute information for all cells in the grid. Instead, incremental update is performed.

Clustering with obstacles is a novel and interesting problem in the fields of data mining. While the work presented here is sufficient for many applications of clustering with obstacles entities, there still a lot of future work to be considered. We belief that the future work may go in several directions. The first direction is to accommodate the existent algorithms like STING[31] that querying spatial databases to consider the presence of obstacles. It will be useful to extend *SCLD* to solve the clustering problem in a database of *D* dimensions. It will also be useful to consider the clustering problem with obstacles when the obstacles are complex shapes rather than simple polygons.

## ملخص الرسالة

تغيرت تطبيقات قواعد البيانات من حيث التعقيد و الصعوبة من تطبيقات حفظ البيانات البسيطة إلى تطبيقات استخدام قواعد البيانات في الاستعانة بالحاسب في التصميم و نظم المعلومات الجغرافية.

لكي تتعامل نظم إدارة قواعد البيانات مع التطبيقات الحديثة لابد من تطوير أداءها ليحقق كفاءة عالية و للتغلب على المشكلات الجديدة آلتي تظهر عند استخدام هذه التطبيقات الحديثة . تقدم الرسالة العديد من الخوارزميات ذات الكفاءة العالية التي تقوم بعمليه التجميع(clustering) في قواعد البيانات المكانية. و تعتبر عمليه التجميع أحد أهم وسائل التنقيب في قواعد البيانات. تهدف عمليه التنقيب في قواعد البيانات إلى استخراج نماذج غير موجدة صراحة في قواعد البيانات.

تقدم الرسالة خوارزم من النوع الشبكي(grid-based) الذي له العديد من المزايا وله أيضا أداء مقارن بأفضل خوارزم تجميع. الخوارزم له العديد من المزايا. الخوارزم لا يطلب العديد من المعملات حول نطاق البيانات. الخوارزم يطلب فقط عدد النقط المطلوب تجميعها و عدد الخلايا في الشبكه. الخوارزم له القدرة علي اكتشاف التجمعات ذات الأشكال المختلفة. درجه تعقيد الخوارزم مقارنه بدرجه تعقيد أفضل خوارزم تجميع. أثبتت النتائج المعملية لتنفيذ الخوارزم علي قواعد بيانات ذات أحجام مختلفة أن وقت تنفيذ الخوارزم أفضل بكثير من وقت تنفيذ واحد من أشهر خوارزميات التجميع(CLARANS). أثبتت النتائج أن أداء الخوارزم لا يسوء مع زيادة عدد النقط المطلوب تجميعها.

المشاركة الثانية للرسالة تعديل الخوارزم المقترح لكي يتعامل مع العقبات(obstacles). معظم الخوارزميات التي تقوم بعمليه التجميع(clustering) في قواعد البيانات المكانية لا تأخذ في الاعتبار وجد العقبات في الطبيعية. هذه العقبات تشمل عقبات طبيعية مثل الأنهار و الجبال أو عقبات صناعية مثل حقول التنقيب. تقدم الرسالة خوارزميان يقومان بعمليه التجميع(clustering) في وجد العقبات في قواعد البيانات المكانية. في البداية قدمنا خوارزم بسيط يتعامل مع العقبات ثم قمنا بتحسين الخوارزم للحصول علي تجمعات(clusters) أكثر دقة. الخوارزميات لها مزايا الخوارزم الشبكي القدم في الرسالة. علاوة علي ذلك تتعامل الخوارزم

مع العقبات ذات الأشكال الهندسية المختلفة. أثبتت النتائج المعملية لتنفيذ الخوارزم الثاني منهم علي قواعد بيانات ذات أحجام مختلفة و في وجد عقبات ذات أشكال هندسية مختلفة أن وقت تنفيذ الخوارزم أفضل بكثير من وقت تنفيذ خوارزم التجميع ) في وجد العقبات (COD-CLARANS). أثبتت النتائج أن أداء الخوارزم يفوق (COD-CLARANS) بصورة عامة وجه الخصوص مع زيادة عدد النقط المطلوب تجميعها.

# أساليب ذات كفاءة عالية للتنقيب في قواعد البيانات المكانية

رسالة ماجستير مقدمة من الطالب

**محمد عبد المنعم محمود محمد الزواوى**

إلى

**قسم الرياضيات – كلية العلوم – جامعة القاهرة**

لاستيفاء الحصول علي درجة الماجستير في علوم الحاسب

يونيو ٢٠٠٢

المشرفون

| | |
|---|---|
| **د. محمد عز الدين الشرقاوى** | **أ.د.ليلى فهمي عبد العال** |
| قسم نظم المعلومات | قسم الرياضيات |
| كلية الحاسبات و المعلومات | كلية العلوم |
| جامعة القاهرة | جامعة القاهرة |